\documentclass[a4paper]{amsart}

\usepackage[utf8]{inputenc}
\usepackage[english]{babel}
\usepackage[T1]{fontenc}
\usepackage{amsmath}
\usepackage{amsthm}
\usepackage{amssymb}
\usepackage{aligned-overset}
\usepackage{xcolor}
\usepackage{forest}
\usepackage{tikz}
\usetikzlibrary{automata}
\usetikzlibrary{positioning}
\usepackage{mathtools}

\usepackage{enumerate}
\usepackage{float}

\usepackage{hyperref}
\theoremstyle{definition}
\newtheorem{defi}{Definition}[section]
\newtheorem{example}[defi]{Example}

\theoremstyle{plain}
\newtheorem{theorem}[defi]{Theorem}

\newtheorem{lemma}[defi]{Lemma}
\newtheorem{cor}[defi]{Corollary}

\begin{document}

\title{Average Case Analysis of Leaf-Centric Binary Tree Sources}

\author{Louisa Seelbach Benkner, Markus Lohrey and Stephan Wagner}
\thanks{A short version of this paper appeared in the Proceedings of MFCS 2018 \cite{SeelbachLo18}.
The first and the second author have been supported by the DFG research project
LO 748/10-1 (QUANT-KOMP). The third author is supported by the Knut and Alice Wallenberg Foundation, grant KAW 2017.0112.}

\begin{abstract} We study the average number of distinct fringe subtrees in random trees generated by leaf-centric binary tree sources as introduced by Zhang, Yang and Kieffer in \cite{ZhangYK14}. A leaf-centric binary tree source induces for every $n \geq 2$ a probability distribution on the set of binary trees with $n$ leaves. We generalize a result by Flajolet, Gourdon and Martinez \cite{FlajoletGM97} and Devroye \cite{Devroye98}, according to which the average number of distinct fringe subtrees in a random binary search tree of size $n$ is in $\Theta(n/\log n)$,  as well as a result by Flajolet, Sipala and Steayert \cite{FlajoletSS90}, according to which the number of distinct fringe subtrees in a uniformly random binary tree of size $n$ is in $\Theta(n/\sqrt{\log n})$.
\end{abstract}

\maketitle

\allowdisplaybreaks

\section{Introduction}

A \emph{fringe subtree} of a rooted tree is a subtree which consists of a node and all its descendants. 
Fringe subtrees are a natural object of study in the context of
random trees, and there is a great variety of results for various random tree models, see e.g. \cite{aldous91,dennertgr10,devroye14,FengM10}.

Fringe subtrees are of particular interest in computer science. A widely used lossless compression method for rooted trees
is to represent a tree as a directed acyclic graph (DAG), which is obtained by merging
nodes that are roots of identical fringe subtrees (see Figure~\ref{fig:dag} for an example). This compressed representation
of the tree is often shortly referred to as minimal DAG and its size (number
of nodes) is the number of distinct fringe subtrees occurring in the tree. Compression by minimal DAGs has found applications in many areas of
computer science, as for example in compiler construction \cite[Chapter 6.1 and 8.5]{AhoSU86}, unification \cite{PatersonW78}, symbolic model checking (binary decision diagrams) \cite{Bry92},
information theory \cite{GanardiHLS19, ZhangYK14} and XML compression and querying \cite{BuGrKo03,FrGrKo03}.

In this work, we consider the problem of deriving asymptotic estimates for the
expected number of distinct fringe subtrees in random binary trees. 
Binary trees will be rooted ordered trees where every node is either a leaf or has
exactly two children (a left child and a right child). We define the size of a binary
tree as the number of leaves.\footnote{This has the technical advantage that for every $n$ there is a binary tree of size $n$.}
So far, the average number of distinct fringe subtrees has mainly been studied with respect to two probability distributions: uniformly random binary trees, and random binary search trees. A uniformly random ordered binary
tree of size $n$ is a random tree whose corresponding probability distribution is
the uniform probability distribution on the set of ordered binary trees of size $n$.
In \cite{FlajoletSS90}, Flajolet, Sipala and Steyaert proved that the expected number of distinct fringe subtrees in a uniformly random ordered binary tree of size $n$ is asymptotically equal to $c\cdot n/\sqrt{\ln n}$, where $c$ is the constant $c=2\sqrt{\ln4/\pi}$. This result of Flajolet et al.~was extended to unranked labeled trees in \cite{MLMN13} (for a different constant $c$). Moreover, an alternative proof to the result of Flajolet et al.~was presented in \cite{RalaivaosaonaW15} in the context of simply generated families of trees.

Another important type of random trees are random binary search trees. A random binary search tree of size $n$ is a binary search tree built by inserting the keys $\{1, \dots, n\}$ according to a uniformly chosen random permutation on $\{1, \dots, n\}$. Random binary search trees naturally arise in theoretical computer science, see e.g. \cite{Drmota09}. In \cite{FlajoletGM97}, Flajolet, Gourdon and Martinez proved that the expected number of distinct fringe subtrees in a random binary search tree of size $n$ is in $\mathcal{O}(n/\log n)$. 
This result was improved in \cite{Devroye98} by Devroye, who showed that the asymptotics $\Theta(n/\log n)$ holds, and reproved in a recent paper by Bodini et al.~\cite{BodiniGGLN20}. The exact multiplicative constant was finally determined by the third author in \cite{Wa24}, who proved
that the number of distinct fringe subtrees in a random binary search tree of size $n$ is $c \cdot n/\ln n \cdot (1+o(1))$ for $c \approx 2.4071298335$
both in expectation and with high probability.

\begin{figure}\label{fig:dag}
\centering
\begin{minipage}{0.45\textwidth}
\centering

\tikzset{level 1/.style={sibling distance=20mm}}
		\tikzset{level 2/.style={sibling distance=15mm}}
		\tikzset{level 3/.style={sibling distance=15mm}}  
		\tikzset{level 4/.style={sibling distance=8.5mm}}
	
\begin{tikzpicture}[scale=0.8,auto,swap,level distance=10mm,  ]
		\node[circle, inner sep = 2pt,fill=black] (eps) {} 
		child {node[circle, fill=black, inner sep = 2pt,minimum size = 1pt]{}
		child {node[circle, fill=black, inner sep = 2pt,minimum size = 1pt]{}
	child {node[circle, fill=black, inner sep = 2pt,minimum size = 1pt]{}
	child {node[circle, fill=black, inner sep = 2pt,minimum size = 1pt]{}}
		child {node[circle, fill=black, inner sep = 2pt,minimum size = 1pt]{}}
	}
		child {node[circle, fill=black, inner sep = 2pt,minimum size = 1pt]{}
	child {node[circle, fill=black, inner sep = 2pt,minimum size = 1pt]{}}
		child {node[circle, fill=black, inner sep = 2pt,minimum size = 1pt]{}}	
		}	
		}
	child {node[circle, fill=black, inner sep = 2pt,minimum size = 1pt]{}}}
		child {node[circle, fill=black, inner sep = 2pt,minimum size = 1pt]{}
	child {node[circle, fill=black, inner sep = 2pt,minimum size = 1pt]{}}
		child {node[circle, fill=black, inner sep = 2pt,minimum size = 1pt]{}
		child {node[circle, fill=black, inner sep = 2pt,minimum size = 1pt]{}}
		child {node[circle, fill=black, inner sep = 2pt,minimum size = 1pt]{}}}	
		}
;
\end{tikzpicture}
\end{minipage}
\begin{minipage}{0.45\textwidth}
\centering
\begin{tikzpicture}[scale=0.8,auto,swap,level distance=10mm ]
		 \node[circle, inner sep = 2pt,fill=black] (A) at (1,4){} ;
    \node[circle, inner sep = 2pt,fill=black] (B) at (0,3){};
    \node[circle, inner sep = 2pt,fill=black](C) at (2,3){};
    \node[circle, inner sep = 2pt,fill=black] (D) at (1,2){} ;
    \node[circle, inner sep = 2pt,fill=black](E) at (-1,2) {};
    \node[circle, inner sep = 2pt,fill=black](F) at (1,0.5) {};

    \draw [-stealth](A) edgenode[right]{} (B);
     \draw [-stealth] (A) edgenode[left]{} (C);
      \draw [-stealth] (B) edge node[right]{} (D);
       \draw [-stealth] (C) edge node[right]{} (D);
    \draw [-stealth](B) edge node[right]{}  (E);
    \draw [-stealth](C) edge[bend left] node[right]{}  (F);
    \draw [-stealth](E) edge[bend left] node[right]{}(F);
    \draw [-stealth](E) edge[bend right] node[right]{}(F);
    \draw [-stealth](F) edge[bend right]  node[left]{} (D);
    \draw [-stealth](F) edge[bend left] node[left]{}(D);
\end{tikzpicture}
\end{minipage}
\caption{A binary tree (left) and its corresponding minimal DAG (right).}
\end{figure}

The goal of this work is to extend the results from \cite{Devroye98,FlajoletGM97, FlajoletSS90} by proving upper and lower bounds for the average number of distinct fringe
subtrees that hold for large classes of distributions. A very general concept to model probability distributions on the set of binary trees of size $n$ was presented by Kieffer et al.~in \cite{KiefferYS09} (see also \cite{ZhangYK14}), where the authors introduce \emph{leaf-centric binary tree sources} as an extension of the classical notion of an information source on finite sequences to binary trees.

A leaf-centric binary tree source is induced by a mapping $\sigma: \mathbb{N}\times\mathbb{N}\to[0,1]$ that satisfies $\sum_{i=1}^{n-1}\sigma(i,n-i)=1$ for every $n \geq 2$. In other words, $\sigma$ restricted to the set of ordered pairs $S_n:=\{(i,n-i) \mid 1 \leq i \leq n-1\} $ is a probability mass function for every $n \geq 2$. 
A leaf-centric binary tree source randomly generates a binary tree of size $n$ as follows: we start with a single root node labeled with $n$ and randomly choose a pair $(i,n-i)$ according to the distribution $\sigma$ on $S_n$. Then, a left (resp. right) child labeled with $i$ (resp. $n-i$) is attached to the root, and the process is repeated with these two nodes. The process stops at nodes with label $1$ (and in the end, the node labels are omitted). In particular, such a mapping $\sigma$ induces a function $P_{\sigma}$ which restricts to a probability mass function on the set of binary trees of size $n$ for every $n \geq 1$.

The binary search tree model is the leaf-centric binary tree source where the mapping $\sigma$ corresponds to the uniform probability distribution on the set $S_n$ for every $n \geq 2$ (Example~\ref{ex:bst}). Moreover, the uniform probability distribution on the set of binary trees of size $n$ can be modeled as a leaf-centric binary tree source as well (Example~\ref{ex:uniform}). Other well-known leaf-centric tree sources are the binomial random tree model \cite{KiefferYS09} (see Example~\ref{ex:dst}), where the mapping $\sigma$ corresponds to a binomial distribution on $S_n$,
and the critical $\beta$-splitting random tree model x\cite{AlJa25,AlPi25,AlJa24arxiv,AlJa25arxiv}; see Example~\ref{ex:beta}.

Let $F_{n,\sigma}$ denote the (random) number of distinct (ordered) fringe subtrees occurring in a random tree of size $n$ generated by the leaf-centric binary tree source corresponding to the mapping $\sigma$. In this work, we investigate the average $\mathbb{E}(F_{n,\sigma})$ under certain conditions on the mapping $\sigma$, that is, we assume that the mapping $\sigma$ satisfies certain properties and then derive upper and lower bounds on $\mathbb{E}(F_{n,\sigma})$ with respect to these conditions.
In particular, we consider four classes of leaf-centric binary tree sources, listed below, for which we will be able to derive asymptotic upper or lower bounds on $\mathbb{E}(F_{n,\sigma})$. Here and in the rest of the paper, increasing/decreasing functions are not necessarily strictly monotone: if $f$ is increasing, then $f(x) \geq f(y)$ whenever $x > y$, but not necessarily $f(x) > f(y)$ (and analogously for decreasing functions).
\begin{itemize}
\item[(i)] A leaf-centric tree source with mapping $\sigma$ is called \emph{$\psi$-upper-bounded} for a decreasing function $\psi: \mathbb{R} \to (0,1]$ if there is a constant $N$ such that
for all $1 \leq i \leq n-1$ and $n \geq N$: 
\begin{align*}
\sigma(i,n-i)\leq \psi(n).
\end{align*}
\item[(ii)] A leaf-centric tree source with mapping $\sigma$ is called \emph{$\phi$-weakly-balanced} for a decreasing function $\phi: \mathbb{N} \to (0,1]$ if there is a constant $\gamma \in (0,\frac{1}{2})$ and an integer $N$ such that for all $n \geq N$:
\begin{align*}
\sum_{\gamma n \leq i \leq (1-\gamma)n}\sigma(i,n-i) \geq \phi(n).
\end{align*}
\item[(iii)] A leaf-centric tree source with mapping $\sigma$ is \emph{$\vartheta$-strongly-balanced} for a decreasing function $\vartheta: \mathbb{R} \to (0,1]$ if there is a constant $\gamma \in (0,\frac{1}{2})$, an integer $N$ and a constant $C$ such that for all $n \geq N$ and $C \leq r \leq \lceil \gamma n\rceil$:
\begin{align*}
\sum_{r \leq i \leq n-r}\sigma(i,n-i)\geq \vartheta(r).
\end{align*}
\item[(iv)] A leaf-centric tree source with mapping $\sigma$ is called \emph{$\xi$-unbalanced} for a decreasing function $\xi$ if there is a constant $\gamma \in (0,\frac{1}{2})$, an integer $N$ and a constant $C$ such that for all $n \geq N$ and $C \leq r \leq \lceil \gamma n \rceil$:
\begin{align*}
\sum_{r \leq i \leq n-r}\sigma(i,n-i) \leq \xi(r).
\end{align*}
\end{itemize}
Property (i) quantifies how close $\sigma$ is to the binary search tree model. The latter is $\psi$-upper-bounded for $\psi(n) = 1/(n-1)$, which is the smallest function $\psi$  for which (i) makes
sense (since  the sum over all $\sigma(i,n-i)$ for $1 \leq i \leq n-1$ has to be one).
Property (ii) generalizes the concept of balanced binary tree sources from \cite{GanardiHLS19,ZhangYK14}: When randomly constructing a binary tree with respect to a leaf-centric tree source of
type (ii), the probability that the current weight is roughly equally split among the two children is bounded below by the function $\phi$. Therefore, for slowly decreasing
functions $\phi$, balanced trees are preferred by this model. Property (iii) is a stronger constraint than property (ii): Every leaf-centric tree source that is $\vartheta$-strongly-balanced for a function $\vartheta$ is also $\phi$-weakly-balanced for the function $\phi$ defined by $\phi(n)=\vartheta(\lceil \gamma n\rceil )$.

As our main results, we obtain the following asymptotic bounds on the expected number of distinct fringe subtrees for these classes of leaf-centric binary tree sources.
Let $\sigma$ be a leaf-centric tree source.
\begin{itemize}
\item[(a)] If $\sigma$ is $\psi$-upper-bounded, then $\mathbb{E}(F_{n,\sigma}) \in \mathcal{O}(n\psi(\log n))$ (Theorem~\ref{thm:upper-bounded-theorem}, where as slightly more general result is shown).
\item[(b)] If $\sigma$ is $\psi$-upper-bounded and there are constants $N \in \mathbb{N}$, $\rho<1$ such that $\psi(x) < \rho$ for all $x \geq N$, then $\mathbb{E}(F_{n,\sigma}) \in \Omega\big(\frac{n}{\log n}\big)$ (Theorem~\ref{thm:lower-bound-information-theoretic}).
\item[(c)] If $\sigma$ is $\phi$-weakly-balanced, then $\mathbb{E}(F_{n,\sigma}) \in \mathcal{O}\big(\frac{n}{\phi(n)\log n}\big)$ (Theorem~\ref{thm:weakly-balanced}).
\item[(d)] If $\sigma$ is $\vartheta$-strongly-balanced, then $\mathbb{E}(F_{n,\sigma}) \in \mathcal{O}\big(\frac{n}{\vartheta(\log n)\log n}\big)$ (Theorem~\ref{thm:strongly-balanced}).
\item[(e)] If $\sigma$ is $\xi$-unbalanced and some additional technical conditions are satisfied, then $\mathbb{E}(F_{n,\sigma}) \in \Omega\big(\frac{n}{\xi(\log n)\log n}\big)$ (Theorem~\ref{thm:lower-bound-unbalanced}).
\end{itemize}
The precise statements and asymptotic bounds including leading constants for the main term are given in the respective theorems. These results generalize the results from \cite{Devroye98, FlajoletGM97,Wa24}  on the binary search tree model in the following sense:
from each of the results (a), (c) and (d) (respectively, (b) and (e)), we obtain the upper bound $\mathcal{O}(n/\log n)$ (respectively, the lower bound $\Omega(n/\log n)$) on the expected number of distinct fringe subtrees occurring in a random binary search tree of size $n$. Furthermore, results (d) and (e) combined yield that the expected number of distinct fringe subtrees in a uniformly random binary tree of size $n$ is in $\Theta(n/\sqrt{\log n})$, as shown in \cite{FlajoletSS90, RalaivaosaonaW15}. Finally, results (b) and (c) imply the asymptotic bound $\Theta(n/\log n)$ on the expected number of distinct fringe subtrees in a random binary tree of size $n$ generated according to the binomial random tree model. 
For the binary search tree model and uniformly random binary trees we do not get the exact multiplicative constants obtained in 
\cite{FlajoletSS90, RalaivaosaonaW15,Wa24}. This is the price that we have to pay for dealing with classes of random tree models instead
of specific models.

The general message of our results could be stated as follows: if a leaf-centric tree source produces 
balanced trees with high (resp., small) probability (which is formalized by the above classes (ii), (iii) and (iv)), then the average number of distinct fringe subtrees gets smaller (resp., higher). 

A short version of this work appeared in the Proceedings of MFCS 2018 \cite{SeelbachLo18}: The results (a), (b) and (c) already appeared in the conference version \cite{SeelbachLo18} (however, among others, several proofs needed for the respective results have been simplified, and leading constants in front of the main terms in the asymptotic bounds have been improved in this version). The results (d) and (e) are new contributions of this long version.

\medskip

\paragraph{\bf Related work}

Recently, in \cite{SeelbachW22}, the problem of estimating the number of distinct fringe subtrees in random trees was investigated in a more general setting, i.e., under a generalized interpretation of ``distinctness'', which allows for many different interpretations of what ``distinct'' fringe subtrees are. While in the aforementioned results from \cite{Devroye98, FlajoletGM97, FlajoletSS90, RalaivaosaonaW15}, two fringe subtrees were considered as ``distinct'' if they are distinct as ordered trees, the generalized interpretation of distinctness from \cite{SeelbachW22} allows for example to count the number of distinct \emph{unordered} fringe subtrees in ordered random trees of size $n$, i.e., two fringe subtrees are considered as ``distinct'' if their corresponding unordered trees are distinct. 
Amongst others, as special cases of more general theorems, it is shown in \cite{SeelbachW22} that the number of distinct unordered fringe subtrees in a uniformly random binary tree of size $n$ is in $\Theta(n/\sqrt{\log n})$ and that the number of distinct unordered fringe subtrees in a random binary search tree of size $n$ is in $\Theta(n/\log n)$, where both asymptotic results hold not only in expectation, but also with high probability, i.e., with probability tending to $1$ as $n \to \infty$ (for the precise upper and lower bounds including leading constants, see \cite{SeelbachW22}). Moreover, the results from \cite{FlajoletSS90, RalaivaosaonaW15} (for the uniform distribution) and \cite{Devroye98, FlajoletGM97} (for the binary search tree model) were generalized in \cite{SeelbachW22} in the sense that these results hold not only in expectation, but also with high probability.

Leaf-centric binary tree sources  have been used in information theory in order
to obtain universal tree coders \cite{GanardiHLS19,KiefferYS09,MunroNSW21,ZhangYK14}. In particular, it has been shown in \cite{ZhangYK14} that for every
leaf-centric tree source $\sigma$ with $\mathbb{E}(F_{n,\sigma}) \in o(n)$, the average-case redundancy of a certain tree encoder converges to zero;
see \cite{ZhangYK14} for precise definitions. Our upper bounds from Theorems~\ref{thm:upper-bounded-theorem}, \ref{thm:weakly-balanced} and \ref{thm:strongly-balanced} yield large classes of leaf-centric tree sources with this property.

\section{Preliminaries}\label{sec:preliminaries}

We use the classical Landau notations $\mathcal{O}$, $o$, $\Omega$ and $\omega$.
In the following, $\log x$ will denote the binary logarithm $\log_2 x$ of a positive real number $x$.
Let $\mathbb{N}$ denote the natural numbers without zero.

\subsection{Binary Trees}

We write $\mathcal{T}$ for the set of \emph{(full) binary trees}, i.e., of rooted ordered trees such that each node has either exactly two or zero children.
The \emph{size} of a binary tree $t$ is the number of leaves of $t$ and denoted by $|t|$. 
We use $\mathcal{T}_n$ to denote the set of binary trees which have exactly $n$ leaves. It is well known that $|\mathcal{T}_n|=C_{n-1}$, where $C_{n}$ denotes the $n$-th \emph{Catalan number} \cite{FlajoletS09}. We have
\begin{align}\label{eq:Catalan-growth}
C_n =\frac{1}{n+1}\binom{2n}{n} \sim \frac{4^n}{\sqrt{\pi}n^{3/2}}(1+\mathcal{O}(1/n)),
\end{align}
where the asymptotic growth follows from Stirling's formula (see e.g. \cite{FlajoletS09}).

A \emph{fringe subtree} of a tree is a subtree that consists of a node and all its descendants.
For a node $v$ of a binary tree $t \in \mathcal{T}$, let $t[v]$ denote the fringe subtree of $t$ which is rooted at $v$. Furthermore, for an inner node $v$ of $t$, let $t_{\ell}[v]$ (resp.~$t_r[v]$) denote the fringe subtree rooted at $v$'s left (resp.~right) child node. 
If $v$ is the root node of $t$, we write $t_{\ell}$ and $t_r$ for $t_{\ell}[v]$ and $t_r[v]$ and call $t_{\ell}$ (resp.~$t_r$) the \emph{left subtree} (resp. \emph{right subtree}) of $t$.

\subsection{Leaf-centric binary tree sources}\label{leafcentric}

Let $\mathcal{L}$ denote the set of all functions $\sigma \colon \mathbb{N} \times \mathbb{N} \rightarrow [0,1]$ that satisfy 
\begin{align}\label{sigmacond}
\sum_{i,j \geq 1 \atop i+j=k} \sigma(i,j) = 1 
\end{align}  
for every integer $k \geq 2$. A mapping $\sigma\in \mathcal{L}$ induces a probability mass function $P_{\sigma}: \mathcal{T}_n \rightarrow [0,1]$ for every $n \geq 1$ in the following way: define $P_{\sigma} \colon \mathcal{T} \rightarrow [0,1]$ inductively by 
\begin{equation} \label{eq-prob-mass-function-P}
P_{\sigma}(t)=\begin{cases}
1 \quad &\text{if } |t| =1,\\
\sigma(|t_{\ell}|,|t_r|) \cdot P_{\sigma}(t_{\ell}) \cdot P_{\sigma}(t_r) \quad &\text{otherwise.}
\end{cases}
\end{equation}
A tuple $(\mathcal{T}, (\mathcal{T}_n)_{n \in \mathbb{N}}, P_{\sigma})$ with $\sigma \in \mathcal{L}$ is called a \emph{leaf-centric tree source} \cite{ZhangYK14, KiefferYS09}.
Furthermore, the following notation will be useful: for a function $\sigma \in \mathcal{L}$, we define $\sigma^*: \mathbb{N} \times \mathbb{N} \to [0,1]$ as
\begin{align}\label{eq:sigma-star}
\sigma^*(i,j)=\begin{cases} \sigma(i,j)+\sigma(j,i) \quad &\text{ if }  i \neq j,\\
\sigma(i,j) &\text{ if } i=j.
\end{cases}
\end{align}
Note that $\sigma^*(i,j)\leq 1$ for all $i,j \in \mathbb{N}$ and that $\sum_{1\le i \le n/2}\sigma^*(i,n-i)=1$ (where the sum goes over all integers in the interval $[1,n/2]$).

Intuitively, a leaf-centric binary tree source randomly generates a binary tree of size $n$ as follows: we start at the root node and determine the sizes of the left and of the right subtree, where the probability that the left subtree is of size $i$ for $i \in \{1,\dots, n-1\}$ (and consequently, the right subtree is of size $n-i$) is given by $\sigma(i,n-i)$. This process then recursively continues in the left and right subtree, i.e., the leaf-centric binary tree source then randomly generates a binary tree of size $i$ as the left subtree and a binary tree of size $n-i$ as the right subtree.
In particular, for $t \in \mathcal{T}$, we have
\begin{align*}
P_{\sigma}(t)=\!\!\!\!\prod_{v \text{ inner node of } t}\sigma(|t_{\ell}[v]|,|t_r[v]|).
\end{align*}

Several well-known random tree models can be characterized as leaf-centric tree sources (see Figure~\ref{fig:probability-trees} for an example of two binary trees together with the probabilities assigned to them by the leaf-centric tree sources from the following examples).

\begin{example}\label{ex:bst}
The \emph{binary search tree model} \cite{Drmota09, FlajoletGM97} corresponds to the leaf-centric binary tree source defined by 
\begin{align*}
\sigma_{\text{bst}}(i,n-i)=\frac{1}{n-1}
\end{align*}
for every $n \geq 2$ and $1 \leq i \leq n-1$. This distribution over binary trees arises if a binary search tree of size
$n$ is built by inserting the keys $1,\dots,n$ according to a uniformly chosen random permutation on $1,\dots, n$ \cite{Drmota09}.
\end{example}

\begin{example}\label{ex:uniform}
Perhaps the most natural probability distribution on the set of binary trees $\mathcal{T}_n$ is the uniform probability distribution, i.e., every binary tree $t \in \mathcal{T}_n$ is assigned the same probability $1/C_{n-1}$. This probability distribution corresponds to the leaf-centric binary tree source with 
\begin{align*}
\sigma_{\text{uni}}(i,n-i)=\frac{C_{i-1}C_{n-i-1}}{C_{n-1}}
\end{align*}
for every $n \geq 2$ and $1 \leq i \leq n-1$ \cite{KiefferYS09}.
\end{example}

\begin{example}\label{ex:dst}
Fix a constant $p \in (0,1)$ and define
\begin{equation} \label{eq-dst}
\sigma_{\text{bin},p}(i,n-i)=p^{i-1}(1-p)^{n-i-1}\binom{n-2}{i-1}
\end{equation}
for every $n\geq 2$ and $1 \leq i \leq n-1$. This leaf-centric binary tree source corresponds to the \emph{binomial random tree model}, which was studied in \cite{KiefferYS09} for the case $p=1/2$, and which is a slight variant of the \emph{digital search tree model}, \cite{Drmota09, Martinez92}.
\end{example}

\begin{example}\label{ex:beta}
The critical $\beta$-splitting random tree model is obtained by taking 
\begin{equation} \label{eq-beta}
\sigma_{\beta}(i,n-i)= \frac{n}{2 h_{n-1} i (n-i)},
\end{equation}
where $h_{n-1}$ is the $(n-1)$-th harmonic number.
It has been investigated in \cite{AlJa25,AlPi25,AlJa24arxiv,AlJa25arxiv}.
\end{example}

\begin{figure}[t]

\tikzset{level 1/.style={sibling distance=10mm}}
		\tikzset{level 2/.style={sibling distance=10mm}}
		\tikzset{level 3/.style={sibling distance=10mm}}  
		\tikzset{level 4/.style={sibling distance=10mm}}

		\centering
	\begin{tikzpicture}[scale=0.8,auto,swap,level distance=10mm,  ]
		\node[circle, inner sep = 2pt,fill=black] (sk1) {} 
		child {node[circle, fill=black, inner sep = 2pt,minimum size = 1pt]{}
		child {node[circle, fill=black, inner sep = 2pt,minimum size = 1pt]{}
		child {node[circle, fill=black, inner sep = 2pt,minimum size = 1pt]{}}
		child {node[circle, fill=black, inner sep = 2pt,minimum size = 1pt]{}}
		}
	child {node[circle, fill=black, inner sep = 2pt,minimum size = 1pt]{}}}
		child {node[circle, fill=black, inner sep = 2pt,minimum size = 1pt]{}}
;
\node[circle, inner sep = 2pt,draw=none, below=0.25cm  of sk1] (eps) {};
\tikzset{level 1/.style={sibling distance=15mm}}
		\tikzset{level 2/.style={sibling distance=10mm}}
		\tikzset{level 3/.style={sibling distance=10mm}}  
		\tikzset{level 4/.style={sibling distance=10mm}}
\node[circle, inner sep = 2pt,fill=black, right=6.5cm  of eps] (eps3) {} 
		child {node[circle, fill=black, inner sep = 2pt,minimum size = 1pt]{}
		child {node[circle, fill=black, inner sep = 2pt,minimum size = 1pt]{}}
	child {node[circle, fill=black, inner sep = 2pt,minimum size = 1pt]{}
	}}
		child {node[circle, fill=black, inner sep = 2pt,minimum size = 1pt]{}
	child {node[circle, fill=black, inner sep = 2pt,minimum size = 1pt]{}}
	child {node[circle, fill=black, inner sep = 2pt,minimum size = 1pt]{}}	
		}
;
\node[draw=none, right=1.1cm of eps](t1){$P_{\sigma_{\text{bst}}}(t_1)=\frac{1}{6}$};
\node[draw=none, below=0.5cm of eps](n1){};
\node[draw=none, right=1.1cm of n1]{$P_{\sigma_{\text{uni}}}(t_1)=\frac{1}{5}$};
\node[draw=none, below=0.5cm of n1](n11){};
\node[draw=none, right=1.1cm of n11]{$P_{\sigma_{\text{bin},1/4}}(t_1)=\frac{1}{64}$};

\node[draw=none, right=1.4cm of eps3](t2){$P_{\sigma_{\text{bst}}}(t_2)=\frac{1}{3}$};
\node[draw=none, below=0.5cm of eps3](n2){};
\node[draw=none, right=1.4cm of n2](n22){$P_{\sigma_{\text{uni}}}(t_2)=\frac{1}{5}$};
\node[draw=none, below=0.5cm of n2](n22){};
\node[draw=none, right=1.4cm of n22]{$P_{\sigma_{\text{bin},1/4}}(t_2)=\frac{3}{8}$};
	\end{tikzpicture}
		\caption{Two binary trees $t_1$ (left) and $t_2$ (right), with the probabilities assigned to them by the binary search tree model (Example~\ref{ex:bst}), the uniform model (Example~\ref{ex:uniform}) and the binomial random tree model (Example~\ref{ex:dst}) with $p=1/4$.}
		\label{fig:probability-trees}
		\end{figure}

In this paper, we are interested in the expected number of \emph{distinct fringe subtrees} in random binary trees (see Figure~\ref{fig:distinct-trees} for an example of a binary tree and its distinct fringe subtrees). Formally, the number of distinct fringe subtrees occurring in a binary tree $t$ can be defined as follows: We define an equivalence relation $\sim$ on the set of nodes of $t$ by $v \sim v'$ if and only if the two fringe subtrees $t[v]$ and $t[v']$ are identical as ordered binary trees. Let $[v]$ denote the equivalence class of node $v$ with respect to this equivalence relation. Then the number of distinct fringe subtrees occurring in $t$ equals the number of equivalence classes $|\{[v] \mid v \text{ node of }t\}|$.
With $T_{n,\sigma}$ we denote a random tree of size $n$ drawn from the set $\mathcal{T}_n$ according to the probability mass function $P_{\sigma}\colon \mathcal{T}_n \to [0,1]$ from
\eqref{eq-prob-mass-function-P}, and with $F_{n,\sigma}$ we denote the (random) number of distinct fringe subtrees occurring in the random tree $T_{n, \sigma}$. 

In the following, we investigate $F_{n,\sigma}$ under certain conditions on the mapping $\sigma$. That is, we will assume that the mapping $\sigma$ satisfies certain properties and then derive upper and lower bounds on $\mathbb{E}(F_{n,\sigma})$. 

In order to obtain upper bounds on $\mathbb{E}(F_{n, \sigma})$, we make use of a technique called \emph{cut-point argument}, which was used before in several works for similar investigations (see for example \cite{Devroye98, FlajoletGM97, RalaivaosaonaW15, SeelbachW22}). The main idea of this cut-point technique is the following: Fix an integer $k >0$.
Let $t \in \mathcal{T}$ be a binary tree. The number of distinct fringe subtrees occurring in $t$ equals 
\begin{enumerate}[(i)]
\item the number of distinct fringe subtrees of size larger than $k$ occurring in $t$ plus 
\item the number of distinct fringe subtrees of size at most $k$ occurring in $t$.
\end{enumerate}
The number (i) of distinct fringe subtrees of size larger than $k$ is then bounded above by the number of \emph{all fringe subtrees} in $t$ of size larger than $k$,\footnote{For instance, the tree in Figure~\ref{fig:distinct-trees} (left) has 7 fringe subtrees of size larger than one, but only 5 different fringe subtrees of size larger than one.}
whereas the number (ii) of distinct fringe subtrees of size at most $k$ is bounded above by the number of \emph{all binary trees} of size at most $k$ (irrespective of their occurrence in $t$). Note that this upper bound on number (ii) is a deterministic quantity; it is a sum of Catalan numbers. Let $Y_{n,k,\sigma}$ denote the (random) number of (all) fringe subtrees of size larger than $k$ occurring in the random tree $T_{n,\sigma}$ (if $\sigma$ is clear from the context, we shortly write $Y_{n,k}$ for $Y_{n,k,\sigma}$). In other words, $Y_{n,k,\sigma}$ is the following random variable:
\begin{equation} \label{eq-Y}
Y_{n,k,\sigma}=|\{v \text{ node of } T_{n,\sigma} \mid |T_{n,\sigma}[v]|> k\}|.
\end{equation}
Note that $Y_{n,k,\sigma}=0$ for all $\sigma \in \mathcal{L}$ if $n \leq k$. Note that the definition of $Y_{n,k,\sigma}$ also makes sense if $k$ is not an integer.
Using the cut-point argument, we obtain the following upper bound on $\mathbb{E}(F_{n,\sigma})$.

\begin{lemma}\label{lemma:upper-bound-Fn}
Let $\sigma \in \mathcal{L}$. The number $F_{n,\sigma}$ of distinct fringe subtrees occurring in the random tree $T_{n,\sigma}$ satisfies
\begin{align*}
\mathbb{E}(F_{n,\sigma}) \leq \mathbb{E}(Y_{n,k,\sigma}) + \sum_{i=0}^{k-1}C_i
\end{align*}
with $Y_{n,k,\sigma}$ as defined in \eqref{eq-Y}.
\end{lemma}
Thus, in Section~\ref{sec:Upper}, we will focus on obtaining upper bounds on $\mathbb{E}(Y_{n,k,\sigma})$ with respect to certain conditions on $\sigma$.
The integer $k$ is called the \emph{cut-point}; it has to be chosen in a suitable way in order to obtain suitable upper bounds on $\mathbb{E}(F_{n,\sigma})$. First, we observe that $\mathbb{E}(Y_{n,k,\sigma})$ satisfies the following recurrence relation.

\begin{figure}[t]

\tikzset{level 1/.style={sibling distance=20mm}}
		\tikzset{level 2/.style={sibling distance=15mm}}
		\tikzset{level 3/.style={sibling distance=15mm}}  
		\tikzset{level 4/.style={sibling distance=8.5mm}}
	
		\centering
		\begin{minipage}{0.45\textwidth}
		\centering
	\begin{tikzpicture}[scale=0.8,auto,swap,level distance=10mm,  ]
		\node[circle, inner sep = 2pt,fill=black] (eps) {} 
		child {node[circle, fill=black, inner sep = 2pt,minimum size = 1pt]{}
		child {node[circle, fill=black, inner sep = 2pt,minimum size = 1pt]{}
	child {node[circle, fill=black, inner sep = 2pt,minimum size = 1pt]{}
	child {node[circle, fill=black, inner sep = 2pt,minimum size = 1pt]{}}
		child {node[circle, fill=black, inner sep = 2pt,minimum size = 1pt]{}}
	}
		child {node[circle, fill=black, inner sep = 2pt,minimum size = 1pt]{}
	child {node[circle, fill=black, inner sep = 2pt,minimum size = 1pt]{}}
		child {node[circle, fill=black, inner sep = 2pt,minimum size = 1pt]{}}	
		}	
		}
	child {node[circle, fill=black, inner sep = 2pt,minimum size = 1pt]{}}}
		child {node[circle, fill=black, inner sep = 2pt,minimum size = 1pt]{}
	child {node[circle, fill=black, inner sep = 2pt,minimum size = 1pt]{}}
		child {node[circle, fill=black, inner sep = 2pt,minimum size = 1pt]{}
		child {node[circle, fill=black, inner sep = 2pt,minimum size = 1pt]{}}
		child {node[circle, fill=black, inner sep = 2pt,minimum size = 1pt]{}}}	
		}
;
\end{tikzpicture}
\end{minipage}
\begin{minipage}{0.5\textwidth}
\centering
\begin{tikzpicture}[scale=0.4,auto,swap,level distance=10mm]
\node[circle, inner sep = 1.5pt,fill=black] (eps2) {} 
		child {node[circle, fill=black, inner sep = 1.5pt,minimum size = 1pt]{}
		child {node[circle, fill=black, inner sep = 1.5pt,minimum size = 1pt]{}
	child {node[circle, fill=black, inner sep = 1.5pt,minimum size = 1pt]{}
	child {node[circle, fill=black, inner sep = 1.5pt,minimum size = 1pt]{}}
		child {node[circle, fill=black, inner sep = 1.5pt,minimum size = 1pt]{}}
	}
		child {node[circle, fill=black, inner sep = 1.5pt,minimum size = 1pt]{}
	child {node[circle, fill=black, inner sep = 1.5pt,minimum size = 1pt]{}}
		child {node[circle, fill=black, inner sep = 1.5pt,minimum size = 1pt]{}}	
		}	
		}
	child {node[circle, fill=black, inner sep = 1.5pt,minimum size = 1pt]{}}}
		child {node[circle, fill=black, inner sep = 1.5pt,minimum size = 1pt]{}
	child {node[circle, fill=black, inner sep = 1.5pt,minimum size = 1pt]{}}
		child {node[circle, fill=black, inner sep = 1.5pt,minimum size = 1pt]{}
		child {node[circle, fill=black, inner sep = 1.5pt,minimum size = 1pt]{}}
		child {node[circle, fill=black, inner sep = 1.5pt,minimum size = 1pt]{}}}	
		}
;
\tikzset{level 1/.style={sibling distance=20mm}}
		\tikzset{level 2/.style={sibling distance=15mm}}
		\tikzset{level 3/.style={sibling distance=8.5mm}}  
		\tikzset{level 4/.style={sibling distance=8.5mm}}
		
\node[circle, inner sep = 1.5pt,fill=black, below=2.5cm of eps2] (eps3) {} 
	child {node[circle, fill=black, inner sep = 1.5pt,minimum size = 1pt]{}
	child {node[circle, fill=black, inner sep = 1.5pt,minimum size = 1pt]{}
	child {node[circle, fill=black, inner sep = 1.5pt,minimum size = 1pt]{}}
		child {node[circle, fill=black, inner sep = 1.5pt,minimum size = 1pt]{}	}}
		child {node[circle, fill=black, inner sep = 1.5pt,minimum size = 1pt]{}	
		child {node[circle, fill=black, inner sep = 1.5pt,minimum size = 1pt]{}}
		child {node[circle, fill=black, inner sep = 1.5pt,minimum size = 1pt]{}	}}
	}
		child {node[circle, fill=black, inner sep = 1.5pt,minimum size = 1pt]{}	}
;
\tikzset{level 1/.style={sibling distance=12mm}}
		\tikzset{level 2/.style={sibling distance=12mm}}
		\tikzset{level 3/.style={sibling distance=8.5mm}}  
		\tikzset{level 4/.style={sibling distance=8.5mm}}
	
\node[circle, inner sep = 1.5pt,fill=black, right=2cm of eps2] (eps4) {} 
	child {node[circle, fill=black, inner sep = 1.5pt,minimum size = 1pt]{}}
	child {node[circle, fill=black, inner sep = 1.5pt,minimum size = 1pt]{}
	child {node[circle, fill=black, inner sep = 1.5pt,minimum size = 1pt]{}}
	child {node[circle, fill=black, inner sep = 1.5pt,minimum size = 1pt]{}}}
;
\tikzset{level 1/.style={sibling distance=20mm}}
		\tikzset{level 2/.style={sibling distance=12mm}}
		\tikzset{level 3/.style={sibling distance=8.5mm}}  
		\tikzset{level 4/.style={sibling distance=8.5mm}}
\node[circle, inner sep = 1.5pt,fill=black, right=2cm of eps3] (eps5) {} 
	child {node[circle, fill=black, inner sep = 1.5pt,minimum size = 1pt]{}
	child {node[circle, fill=black, inner sep = 1.5pt,minimum size = 1pt]{}}
	child {node[circle, fill=black, inner sep = 1.5pt,minimum size = 1pt]{}}}
	child {node[circle, fill=black, inner sep = 1.5pt,minimum size = 1pt]{}
	child {node[circle, fill=black, inner sep = 1.5pt,minimum size = 1pt]{}}
	child {node[circle, fill=black, inner sep = 1.5pt,minimum size = 1pt]{}}}
;
\node[circle, inner sep = 1.5pt,fill=black, right=1.5cm of eps4] (eps6) {} 
;
\tikzset{level 1/.style={sibling distance=15mm}}
		\tikzset{level 2/.style={sibling distance=15mm}}
		\tikzset{level 3/.style={sibling distance=15mm}}  
		\tikzset{level 4/.style={sibling distance=8.5mm}}
	
\node[circle, inner sep = 1.5pt,fill=black, below=2.5cm of eps6] (eps8) {} 
child {node[circle, fill=black, inner sep = 1.5pt,minimum size = 1pt]{}}
	child {node[circle, fill=black, inner sep = 1.5pt,minimum size = 1pt]{}}
;
		\end{tikzpicture}
		\end{minipage}
		\caption{A binary tree (left) and its six distinct fringe subtrees (right).}
		\label{fig:distinct-trees}
		\end{figure}

\begin{lemma}\label{lemma:recurrence-expectation}
Let $\sigma \in \mathcal{L}$, and let $n>k \geq 0$. Then the number $Y_{n,k}$ of fringe subtrees of size larger than $k$ in the random tree $T_{n,\sigma}$ satisfies
\begin{equation} \label{eq-recurrence-expectation1}
\mathbb{E}(Y_{n,k})=1+\sum_{1 \leq i \leq n/2}\sigma^*(i,n-i)\left(\mathbb{E}(Y_{i,k}) +\mathbb{E}(Y_{n-i,k})\right).
\end{equation}
Moreover, if $k+1 > n/2$ we have
\begin{equation} \label{eq-recurrence-expectation2}
\mathbb{E}(Y_{n,k})=1+\sum_{i=k+1}^{n-1}\sigma^*(i,n-i)\mathbb{E}(Y_{i,k}),
\end{equation} 
and if $k+1 \le n/2$ we have
\begin{equation} \label{eq-recurrence-expectation3}
\mathbb{E}(Y_{n,k})=1+\sum_{i=k+1}^{n-k-1}\sigma(i,n-i)\left(\mathbb{E}(Y_{i,k})+\mathbb{E}(Y_{n-i,k})\right) + \sum_{i=n-k}^{n-1}\sigma^*(i,n-i)\mathbb{E}(Y_{i,k}).
\end{equation}
\end{lemma}

\begin{proof}
Let $t \in \mathcal{T}_n$ be a binary tree. Then the number of fringe subtrees of size larger than $k$ occurring in $t$ equals the number of fringe subtrees of size larger than $k$ occurring in its left subtree $t_{\ell}$ plus the number of fringe subtrees of size larger than $k$ occurring in the right subtree $t_r$ plus one (for the tree itself, i.e., the fringe subtree rooted in the root node of $t$). As the left and right subtree of a random tree $T_{n,\sigma}$ conditioned on their sizes $i$ and $n-i$ for some integer $1 \leq i \leq n-1$ 
are again independent random trees $T_{i,\sigma}$ and $T_{n-i,\sigma}$, and as the probability that the left and right subtree are of sizes $i$ and $n-i$ is given by $\sigma(i,n-i)$, we find
\begin{align*}
\mathbb{E}(Y_{n,k})=1+\sum_{i=1}^{n-1}\sigma(i,n-i)\left(\mathbb{E}(Y_{i,k})+\mathbb{E}(Y_{n-i,k})\right).
\end{align*}
From this and the definition of $\sigma^*$ (see \eqref{eq:sigma-star}) we obtain \eqref{eq-recurrence-expectation1}.
With $\mathbb{E}(Y_{i,k})=0$ for $i \leq k$ we can write \eqref{eq-recurrence-expectation1} as
\begin{equation*}
\mathbb{E}(Y_{n,k})=1+\sum_{k+1 \leq i \leq n/2} \!\!\!\! \sigma^*(i,n-i) \mathbb{E}(Y_{i,k}) 
+ \sum_{1 \leq i \leq \min\{n-(k+1),n/2\}} \!\!\!\!\!\!\!\!\!\!\!\!\!\!\!\! \sigma^*(i,n-i)\mathbb{E}(Y_{n-i,k}).
\end{equation*}
By considering the case $k+1 > n/2$ and $k+1 \le n/2$, we finally obtain
\eqref{eq-recurrence-expectation2} and \eqref{eq-recurrence-expectation3}.
\end{proof}

Several times we have to bound expressions of the form $\sum_{1 \le i \le n/2} \sigma^*(i,n-i) \cdot \min \{ \alpha i, \beta \}$
for constants $\alpha, \beta > 0$.
It turns out to be convenient to use the (cumulative) distribution function $D_{\sigma,n} : \mathbb{R}_{\geq 0} \to \mathbb{R}$ 
corresponding to $\sigma^*$, which is defined as follows:
\begin{equation} \label{def-distr-function}
D_{\sigma,n}(x) = \sum_{i \leq x} \sigma^*(i,n-i).
\end{equation}
The sum goes over all integers $i$ with $1 \le i \le \min\{n/2,x\}$. 
If $\sigma$ is clear from the context, we also write $D_n$ for $D_{\sigma,n}$.
In the following lemma we use a Riemann--Stieltjes integral; see e.g. \cite[Chapter~6]{Rudin53}.

\begin{lemma} \label{lemma-integral}
Let $\sigma \in \mathcal{L}$ and $\alpha, \beta > 0$.
Then we have
$$
\sum_{1 \le i \le n/2} \sigma^*(i,n-i) \cdot \min \{ \alpha i, \beta \} = \beta - \int_0^{\beta/\alpha} \alpha D_n(x) \mathrm{d}x .
$$
\end{lemma}

\begin{proof}
Using integration by parts for Riemann--Stieltjes integrals (see e.g.~\cite[p.~141]{Rudin53}), we obtain
\begin{align*}
\sum_{1 \le i \le n/2} \sigma^*(i,n-i) \cdot \min\{ \alpha i, \beta \} & = \int_0^{\infty} \min\{ \alpha x, \beta \} \;\mathrm{d}D_n(x) \\
& = \int_0^{\beta/\alpha} \alpha x \;\mathrm{d}D_n(x) + \int_{\beta/\alpha}^{\infty}  \beta \; \mathrm{d}D_n(x) \\
& = \bigg[ \alpha x D_n(x) \bigg]_0^{\beta/\alpha} \!\!\!\!\!\!- \int_0^{\beta/\alpha}\!\!\! \alpha D_n(x)\; \mathrm{d} x  + \bigg[ \beta D_n(x) \bigg]_{\beta/\alpha}^{\infty} \\
& =  \beta - \int_0^{\beta/\alpha} \alpha D_n(x) \mathrm{d}x .
\end{align*}
This proves the lemma.
\end{proof}

\section{Upper Bounds}\label{sec:Upper}

In this section of the paper, we focus on upper bounds on the expected number $\mathbb{E}(F_{n,\sigma})$ of distinct fringe subtrees occurring in a random tree $T_{n,\sigma}$. In particular, we  present three classes of leaf-centric tree sources, for which we will be able to derive asymptotic upper bounds on $\mathbb{E}(F_{n,\sigma})$.

\subsection{Upper-bounded sources}

The first natural class of leaf-centric tree sources we consider is the following class, where the mapping $\sigma$ (resp. $\sigma^*$) is bounded from above by a function $\psi$:
\begin{defi}[$\psi$-upper-bounded sources]\label{def:upper-bounded}
For a decreasing function $\psi\colon \mathbb{R} \to (0,1]$, let $\mathcal{L}_{\text{up}}(\psi)\subseteq \mathcal{L}$ denote the set of mappings $\sigma \in \mathcal{L}$ for which there is an integer $N_{\sigma}$ such that 
\begin{align*}
\sigma(i,n-i) \leq \psi(n)
\end{align*}
for all $n \geq N_{\sigma}$ and all integers $1 \leq i \leq n-1$. In the same way, let $\mathcal{L}_{\text{up}}^*(\psi)\subseteq \mathcal{L}$ denote the set of mappings $\sigma \in \mathcal{L}$ for which there is an integer $N_{\sigma}$ such that
\begin{align*}
\sigma^*(i,n-i) \leq \psi(n)
\end{align*}
for all $n \geq N_{\sigma}$ and all integers $1 \leq i \leq n-1$. 
\end{defi}
Note that by definition of $\sigma^*$ (see \eqref{eq:sigma-star}), we have $\mathcal{L}_{\text{up}}^*(\psi)\subseteq \mathcal{L}_{\text{up}}(\psi) \subseteq \mathcal{L}_{\text{up}}^*(2\psi)$. Furthermore, note that without loss of generality, we can assume that $\psi(n) \geq 1/(n-1)$ for every integer $n \geq 2$, as the values $\sigma(i,n-i)$ have to add up to $1$ for $1 \leq i \leq n-1$ by condition \eqref{sigmacond}. In the same way, if we consider the class $\mathcal{L}_{\text{up}}^*(\psi)$, we can assume that $\psi(n) \geq 2/n$ for every integer $n \geq 2$.
The class of $\psi$-upper-bounded functions can be generalized as follows:

\begin{defi}[$\psi$-weakly-upper-bounded sources]\label{def:weakly-upper-bounded}
For a decreasing function $\psi\colon \mathbb{R} \to (0,1]$, let $\mathcal{L}_{\text{wup}}(\psi)\subseteq \mathcal{L}$ denote the set of mappings $\sigma \in \mathcal{L}$ for which there is an integer $N_{\sigma}$ such that 
\begin{align*}
\sum_{1 \leq i \leq k}\sigma^*(i,n-i) \leq k \psi(n)
\end{align*}
for all $n \geq N_{\sigma}$ and all integers $1 \leq k \leq n/2$.
\end{defi}
Note that we have $\mathcal{L}_{\text{up}}^*(\psi)\subseteq \mathcal{L}_{\text{wup}}(\psi)$ and hence also $\mathcal{L}_{\text{up}}(\psi)\subseteq \mathcal{L}_{\text{wup}}(2\psi)$.
For $\psi$-weakly-upper-bounded sources we can assume again that $\psi(n) \geq 2/n$ for every integer $n \geq 2$, as the values $\sigma^*(i,n-i)$ have to add up to $1$ for $1 \leq i \leq n/2$ by condition \eqref{sigmacond}.
As our first main theorem, we obtain the following upper bound on $\mathbb{E}(F_{n,\sigma})$ for mappings $\sigma \in \mathcal{L}_{\text{wup}}(\psi)$:

\begin{theorem}\label{thm:upper-bounded-theorem}
Let $\psi\colon \mathbb{R} \to (0,1]$ be a decreasing function, and let $\sigma \in \mathcal{L}_{\text{wup}}(\psi)$. The number $F_{n,\sigma}$ of distinct fringe subtrees in the random tree $T_{n,\sigma}$ satisfies
\begin{align*}
\mathbb{E}(F_{n,\sigma}) \leq 2 n \psi(\log_4 n)+\mathcal{O}\left(\frac{n}{(\log n)^{3/2}}\right).
\end{align*}
\end{theorem}
As $\mathcal{L}_{\text{up}}(\psi) \subseteq \mathcal{L}_{\text{up}}^*(2\psi)$ and $\mathcal{L}_{\text{up}}^*(\psi) \subseteq \mathcal{L}_{\text{wup}}(\psi)$, we immediately obtain an asymptotic upper bound on $\mathbb{E}(F_{n,\sigma})$ for $\sigma \in \mathcal{L}_{\text{up}}(\psi)$ and  $\sigma \in \mathcal{L}_{\text{up}}^*(\psi)$ from Theorem~\ref{thm:upper-bounded-theorem} as well.

In order to prove Theorem~\ref{thm:upper-bounded-theorem}, we make use of the cut-point technique as described in Section~\ref{sec:preliminaries} (see Lemma~\ref{lemma:upper-bound-Fn}). For this, we start with an upper bound on the expectation $\mathbb{E}(Y_{n,k,\sigma})$ of fringe subtrees of size larger than $k$ in a random tree $T_{n,\sigma}$, where $\sigma \in \mathcal{L}_{\text{wup}}(\psi)$ for some mapping $\psi$. 

 \begin{lemma}\label{lemma:upper-bounded}
 Let $\psi: \mathbb{R} \to (0,1]$ be a decreasing function, and let $\sigma \in \mathcal{L}_{\text{wup}}(\psi)$ with integer $N_{\sigma}$. Then the (random) number $Y_{n,k}$ of fringe subtrees of size larger than $k$ in the random tree $T_{n,\sigma}$ satisfies
 \begin{align*}
 \mathbb{E}(Y_{n,k})\leq 2n\psi(k)-2
 \end{align*}
 for all $n,k$ with $n> k\geq N_{\sigma}$.
 \end{lemma}

\begin{proof}
As $\sigma \in \mathcal{L}_{\text{wup}}(\psi)$, we assume that $1 \geq \psi(n) \geq 2/n$ for every integer $n \geq N_{\sigma}$. 
For the distribution function $D_n = D_{\sigma,n}$ in~\eqref{def-distr-function}, we have
\begin{equation} \label{eq-D_n-up}
D_n(x) \leq x\psi(n)
\end{equation}
for all $x \geq 0$.
We prove the statement by induction on $n > k \geq N_{\sigma}$. For the base case, let $n=k+1$. As there is exactly one fringe subtree of size larger than $k$ in a binary tree of size $k+1$ (the fringe subtree rooted at the root node), we have
\begin{align*}
\mathbb{E}(Y_{k+1,k})=1 \leq 2(k+1)\psi(k)-2,
\end{align*}
as $\psi(k)\geq 2/k$ by assumption.
For the induction step, take an integer $n > k+1 >N_{\sigma}$, so that $\mathbb{E}(Y_{i,k})\leq 2i\psi(k)-2$ for every $k < i \leq n-1$. 
Lemma~\ref{lemma:recurrence-expectation} yields
\begin{align*}
\mathbb{E}(Y_{n,k})= 1+\sum _{1 \leq i \leq n/2}\sigma^*(i,n-i)\left(\mathbb{E}(Y_{i,k})+\mathbb{E}(Y_{n-i,k})\right).
\end{align*}
Note that $\mathbb{E}(Y_{i,k}) \leq \max \{2i\psi(k)-2,0\}$ for all $1 \leq i \leq n/2$: if $i>k$, $\mathbb{E}(Y_{i,k}) \leq  2i\psi(k)-2$ holds by the induction hypothesis, and otherwise, if $i \leq k$, we trivially have $\mathbb{E}(Y_{i,k})=0$.
Furthermore, we have $\mathbb{E}(Y_{n-i,k}) \leq 2(n-i)\psi(k)-2$ for all $1 \leq i \leq n/2$:
if $n-i>k$, this holds by the induction hypothesis, and if $n-i\leq k$, we find $\mathbb{E}(Y_{n-i,k})=0<\frac{2n}{k}-2\leq n\psi(k)-2\leq 2(n-i)\psi(k)-2$. We can combine these to
\begin{align*}
\mathbb{E}(Y_{i,k}) + \mathbb{E}(Y_{n-i,k}) &\leq \max \{2i\psi(k)-2,0\} + 2(n-i)\psi(k)-2 \\
&= 2n\psi(k)-2-\min\{2i\psi(k),2\}.
\end{align*}
Thus,  we obtain
\begin{align*}
\mathbb{E}(Y_{n,k}) & \leq 1+\sum _{1 \leq i \leq n/2}\sigma^*(i,n-i)(2n\psi(k)-2-\min\{2i\psi(k),2\})\\
& =  2n\psi(k)-1-\sum _{1 \leq i \leq n/2}\sigma^*(i,n-i)\min\{2i\psi(k),2\} \\
\overset{\text{Lem.~\ref{lemma-integral}}}&{=}  2n\psi(k)- 3 + \int_0^{1/\psi(k)} 2\psi(k) D_n(x) \, \mathrm{d}x  \\
\overset{\text{\eqref{eq-D_n-up}}}&{\leq}  2n\psi(k)-3 + \int_0^{1/\psi(k)} 2\psi(k) \psi(n)x \,\mathrm{d}x  \\
& = 2n\psi(k)-3 + \bigg[ \psi(k)\psi(n) x^2 \bigg]_0^{1/\psi(k)} \\
& = 2n\psi(k)-3 + \frac{\psi(n)}{\psi(k)} \\
& \leq 2n\psi(k)-2,
\end{align*}
where we use the fact that
$\psi$ is decreasing and $n > k$ for the last inequality. 
\end{proof}

We are now able to prove Theorem~\ref{thm:upper-bounded-theorem}:

\begin{proof}[Proof of Theorem~\ref{thm:upper-bounded-theorem}]
Let $\sigma \in \mathcal{L}_{\text{wup}}(\psi)$ with integer $N_{\sigma}$. Let $n > 4^{N_{\sigma}}$, and set $k :=\lceil \log_4 n\rceil > N_{\sigma}$.
We use the cut-point argument from Section~\ref{sec:preliminaries} (see Lemma~\ref{lemma:upper-bound-Fn}) and find that the expected number $\mathbb{E}(F_{n,\sigma})$ of distinct fringe subtrees in the random tree $T_{n,\sigma}$ is bounded above by the expected number $\mathbb{E}(Y_{n,k})$ of fringe subtrees of size larger than $k$ in $T_{n,\sigma}$, plus the number of all binary trees of size at most $k$:
\begin{align*}
\mathbb{E}(F_{n,\sigma}) \leq \mathbb{E}(Y_{n,k})+\sum_{i=0}^{k-1}C_i.
\end{align*}
By Lemma~\ref{lemma:upper-bounded}, we have
\begin{align*}
\mathbb{E}(F_{n,\sigma}) \leq 2n\psi(k)+\sum_{i=0}^{k-1}C_i.
\end{align*}
With the asymptotic growth of the Catalan numbers \eqref{eq:Catalan-growth}, we obtain
\begin{align*}
\mathbb{E}(F_{n,\sigma}) &\leq 2n\psi(k)+\mathcal{O}\left(\frac{4^k}{k^{3/2}}\right)\\
&\leq 2n\psi(\log_4 n)+\mathcal{O}\left(\frac{n}{(\log n)^{3/2}}\right),
\end{align*}
 as $k =\lceil \log_4 n\rceil$.
This yields the asymptotic upper bound from Theorem~\ref{thm:upper-bounded-theorem}.
\end{proof}

Using Theorem~\ref{thm:upper-bounded-theorem}, we obtain an upper bound on the expected number of distinct fringe subtrees in random binary search trees:

\begin{example}\label{ex:upperbounded-bst}
For the binary search tree model from Example~\ref{ex:bst}, we find that $\sigma_{\text{bst}} \in \mathcal{L}_{\text{up}}^*(\psi_{\text{bst}})$, where $\psi_{\text{bst}}\colon \mathbb{R} \to (0,1]$ is given by $\psi_{\text{bst}}(x)=2/(x-1)$. Theorem~\ref{thm:upper-bounded-theorem} gives us the following upper bound on the expected number of distinct fringe subtrees in a random binary search tree:
\begin{align*}
\mathbb{E}(F_{n,\sigma_{\text{bst}}})\leq \frac{8n}{\log n}\left(1+o(1)\right).
\end{align*}
In \cite{Wa24}, it is shown that 
\begin{equation} \label{eq-upper-lower-BST-W24}
 \mathbb{E}(F_{n,\sigma_{\text{bst}}}) = \frac{c \cdot n}{\ln n}  (1+o(1)),
\end{equation} 
where $c\approx 2.4071298335$. Thus, except for the leading constant, the upper bound from Theorem~\ref{thm:upper-bounded-theorem} yields the correct asymptotic growth of $\mathbb{E}(F_{n,\sigma_{\text{bst}}})$.
\end{example}

For the binomial random tree model, we are also able to obtain a non-trivial upper bound on $\mathbb{E}(F_{n,\sigma_{\text{bin},p}})$ from Theorem~\ref{thm:upper-bounded-theorem}:

\begin{example}\label{ex:dst-upper-bounded}
For the binomial random tree model from Example~\ref{ex:dst}, a short computation shows that for a fixed integer $n$, the maximal value of $\sigma_{\text{bin},p}(i,n-i)$ with $1 \leq i \leq n-1$ is attained at $i_{\max}(n) = \lfloor p(n-1) \rfloor +1$. In particular, we have
\begin{align*}
\sigma_{\text{bin},p}(i_{\max}(n), n-i_{\max}(n)) \in \Theta(1/\sqrt{n}).
\end{align*}
Thus, we find that $\sigma_{\text{bin},p} \in \mathcal{L}_{\text{up}}^*(\psi_{\text{bin},p})$ for a function $\psi_{\text{bin},p}$ with $\psi_{\text{bin},p}(x)\in \Theta(1/\sqrt{x})$. By Theorem~\ref{thm:upper-bounded-theorem}, we obtain the bound
$\mathbb{E}(F_{n,\sigma_{\text{bin},p}})\in \mathcal{O}(n/\sqrt{\log n})$. However, in the following sections, we will be able to prove the stronger upper bound $\mathbb{E}(F_{n,\sigma_{\text{bin},p}})\in \mathcal{O}(n/\log n)$ (see Example~\ref{ex:dst-weaklybalanced}) and also a corresponding lower bound $\mathbb{E}(F_{n,\sigma_{\text{bin},p}})\in \Omega(n/\log n)$ (see Example~\ref{ex:dst-information-theoretic}).
\end{example}

\begin{example}\label{ex:upperbounded-beta}
For the critical $\beta$-splitting random tree model (Example~\ref{ex:beta}) we have 
 $\sigma_\beta \in \mathcal{L}_{\text{up}}^*(\psi_\beta)$ with 
 $\psi_\beta(x)= (1+o(1))/\ln(x)$. For this, we use the fact that $h_{n-1} \geq \ln n$ for the $n$-th harmonic number.
  Theorem~\ref{thm:upper-bounded-theorem} gives us the following upper bound on the expected number of distinct fringe subtrees in the
  $\beta$-splitting random tree model:
\begin{align*}
\mathbb{E}(F_{n,\sigma_\beta})\leq \frac{2n}{\ln \log n}\left(1+o(1)\right).
\end{align*}
We are not aware of a lower bound of the same order. Only a lower bound of the form $\Omega(n/\log n)$ will be deduced
in Section~\ref{sec:information-theoretic}.
\end{example}
There are plenty of other ways to choose the mapping $\psi$ in order to obtain non-trivial upper bounds on $\mathbb{E}(F_{n,\sigma})$ for $\sigma \in \mathcal{L}_{\text{up}}^*(\psi)$: For example, $\psi(x)\in \Theta(x^{-\alpha})$ for a constant $0<\alpha \leq 1$ yields $\mathbb{E}(F_{n,\sigma})\in \mathcal{O}(n/(\log n)^\alpha)$. 
Note that Theorem~\ref{thm:upper-bounded-theorem} only makes a non-trivial statement if $\psi(n)<1/2$ for $n \geq N_{\sigma}$. 
For the uniform probability distribution (see Example~\ref{ex:uniform}), Theorem~\ref{thm:upper-bounded-theorem} only yields an upper bound  of the form $\mathbb{E}(F_{n,\sigma_{\text{uni}}})\in \mathcal{O}(n)$, as we have $\sigma_{\text{uni}}(1,n-1) > 1/4$ for every $n \geq 2$.

\subsection{Weakly-balanced sources}\label{sec:weaklybalanced}
In this subsection, we investigate another class of leaf-centric binary tree sources. We consider \emph{weakly-balanced} binary tree sources, which represent a generalization of balanced binary tree sources introduced in \cite{ZhangYK14} and further analyzed in \cite{GanardiHLS19}. 
\begin{defi}[$\phi$-weakly-balanced sources]\label{weakly-balanced}
For a decreasing function $\phi: \mathbb{N} \to (0,1]$ and $\gamma \in (0,\frac{1}{2})$, let $\mathcal{L}_{\text{wbal}}(\phi, \gamma) \subseteq \mathcal{L}$ denote the set of mappings $\sigma\in \mathcal{L}$ for which there is an integer $N_{\sigma}$ such that
\begin{align*}
\sum_{\gamma n \leq i \leq  (1-\gamma)n}\sigma(i,n-i)\geq \phi(n)
\end{align*}
for all $n \geq N_{\sigma}$. 
\end{defi}
For the class of weakly-balanced leaf-centric tree sources, we obtain the following main result:

\begin{theorem}\label{thm:weakly-balanced}
Let $\phi:\mathbb{N}\to (0,1]$ be a decreasing function, let $\gamma \in (0,\frac{1}{2})$, and let $\sigma \in \mathcal{L}_{\text{wbal}}(\phi, \gamma)$. The number $F_{n,\sigma}$ of distinct fringe subtrees in the random tree $T_{n,\sigma}$ satisfies
\begin{align*}
\mathbb{E}(F_{n,\sigma})\leq \frac{n}{\gamma \phi(n)\log_4 n} + \mathcal{O}\left(\frac{n}{(\log n)^{3/2}}\right).
\end{align*}
\end{theorem}
In order to get a nontrivial statement from Theorem~\ref{thm:weakly-balanced}, we should have $\phi(n)\in \Omega(1/\log n)$.
Again, in order to prove Theorem~\ref{thm:weakly-balanced}, we make use of the cut-point argument from Section~\ref{sec:preliminaries} (see Lemma~\ref{lemma:upper-bound-Fn}). We start with the following lemma:

\begin{lemma}\label{lemma:weakly-balanced}
Let $\phi: \mathbb{N} \to (0,1]$ be a decreasing function, let $\gamma \in (0,\frac{1}{2})$, and let $\sigma \in \mathcal{L}_{\text{wbal}}(\phi, \gamma)$ with integer $N_{\sigma}$. Then for $k \geq N_{\sigma}$ and $n \geq k+1$, the (random) number $Y_{n,k}$ of fringe subtrees of size larger than $k$ in the random tree $T_{n, \sigma}$ satisfies
\begin{align*}
\mathbb{E}(Y_{n,k}) \leq \frac{n}{\gamma\phi(n)k}-\frac{1}{\phi(n)}.
\end{align*}
\end{lemma}
\begin{proof}
We prove the statement by induction on $n \geq k+1$. For the base case, let $n=k+1$. As there is exactly one fringe subtree of size larger than $k$ in a binary tree of size $k+1$ (the fringe subtree rooted at the root node), we have
\begin{align*}
\mathbb{E}(Y_{k+1,k})=1 \leq \frac{(\frac{1}{\gamma} -1)(k+1)}{\phi(k+1)k}\leq \frac{k+1}{\gamma\phi(k+1)k}-\frac{1}{\phi(k+1)},
\end{align*}
as $\frac{1}{\gamma}>2$ and $\phi(k+1)\leq 1$ by assumption.
For the induction step, take an integer $n > k+1$, so that 
\begin{equation} \label{eq-ind-hyp-Y_i,k}
\mathbb{E}(Y_{i,k}) \leq \frac{i}{\gamma\phi(i)k}-\frac{1}{\phi(i)}
\end{equation}
for every integer $k+1 \leq i \leq n-1$.
As $\gamma k< \gamma n$ and $\sigma \in \mathcal{L}_{\text{wbal}}(\phi, \gamma)$, we have 
\begin{equation} \label{ineq-D_n-wbal}
D_n(x) \leq 1-\phi(n)
\end{equation}
for every $0 \leq x \leq \gamma k$. 
Lemma~\ref{lemma:recurrence-expectation} yields
\begin{align*}
\mathbb{E}(Y_{n,k})= 1+\sum _{1 \leq i \leq n/2}\sigma^*(i,n-i)\left(\mathbb{E}(Y_{i,k})+\mathbb{E}(Y_{n-i,k})\right).
\end{align*}
First, we observe that for all $1 \leq i \leq n/2$, we have
\begin{align*}
\mathbb{E}(Y_{i,k})\leq \max\left\{\frac{i}{\gamma \phi(n) k}-\frac{1}{\phi(n)},0\right\}.
\end{align*}
To see this, note that (i) $\mathbb{E}(Y_{i,k})=0$ for $i \leq k$, and (ii) \eqref{eq-ind-hyp-Y_i,k} holds for $i > k$ and  
$\phi$ is decreasing.
Furthermore, for all $1 \leq i \leq n/2$ we have
\begin{align*}
\mathbb{E}(Y_{n-i,k})\leq \frac{n-i}{\gamma \phi(n)k}-\frac{1}{\phi(n)}.
\end{align*}
If $n-i>k$, this follows from \eqref{eq-ind-hyp-Y_i,k} and the fact that
$\phi$ is decreasing.
On the other hand, if $n-i \leq k$, then we have
$$\mathbb{E}(Y_{n-i,k})=0<\frac{n-2\gamma k}{2\gamma \phi(n)k}= \frac{n}{2\gamma \phi(n) k}-\frac{1}{\phi(n)} \leq \frac{n-i}{\gamma \phi(n) k}-\frac{1}{\phi(n)},$$ 
as $\gamma < 1/2$ by assumption. Thus, we obtain
\begin{align*}
\mathbb{E}(Y_{n,k})&\leq 1+\sum_{1 \leq i \leq n/2}\sigma^*(i,n-i)\left(\max\left\{\frac{i}{\gamma \phi(n) k}-\frac{1}{\phi(n)},0\right\} + \frac{n-i}{\gamma \phi(n)k}-\frac{1}{\phi(n)}\right)\\
&= 1+\sum_{1 \leq i \leq n/2}\sigma^*(i,n-i)\left(\frac{n}{\gamma \phi(n)k}-\frac{1}{\phi(n)}-\min\left\{\frac{i}{\gamma \phi(n)k},\frac{1}{\phi(n)}\right\}\right)\\
&= 1+\frac{n}{\gamma \phi(n)k}-\frac{1}{\phi(n)}-\sum_{1 \leq i \leq n/2}\sigma^*(i,n-i)\min\left\{\frac{i}{\gamma \phi(n)k},\frac{1}{\phi(n)}\right\} \\
\overset{\text{Lem.~\ref{lemma-integral}}}&{=} 1+\frac{n}{\gamma \phi(n)k}-\frac{2}{\phi(n)}+\int_0^{\gamma k}\frac{D_n(x)}{\gamma \phi(n)k}  \, \mathrm{d}x \\
\overset{\text{\eqref{ineq-D_n-wbal}}}&{\leq} 1+\frac{n}{\gamma \phi(n)k}-\frac{2}{\phi(n)}+\int_0^{\gamma k}\frac{1-\phi(n)}{\gamma \phi(n)k} \, \mathrm{d}x  \\
&= \frac{n}{\gamma \phi(n)k}-\frac{1}{\phi(n)},
\end{align*}
which concludes the proof.
\end{proof}

With Lemma~\ref{lemma:weakly-balanced}, we are now able to prove Theorem~\ref{thm:weakly-balanced}.

\begin{proof}[Proof of Theorem~\ref{thm:weakly-balanced}]
Let $\sigma \in \mathcal{L}_{\text{wbal}}(\psi)$ with integer $N_{\sigma}$. Let $n > 4^{N_{\sigma}}$, and set $k :=\lceil \log_4 n\rceil > N_{\sigma}$.
Again, we use the cut-point technique from Section~\ref{sec:preliminaries} (Lemma~\ref{lemma:upper-bound-Fn}). The expected number $\mathbb{E}(F_{n,\sigma})$ of distinct fringe subtrees in the random tree $T_{n,\sigma}$ is bounded from above by $\mathbb{E}(Y_{n,k})$ plus the number of all binary trees of size at most $k$:
\begin{align*}
\mathbb{E}(F_{n,\sigma}) \leq \mathbb{E}(Y_{n,k}) + \sum_{i=0}^{k-1} C_i.
\end{align*}
With Lemma~\ref{lemma:weakly-balanced}, and with the asymptotic growth of the Catalan numbers, we find 
\begin{align*}
\mathbb{E}(F_{n,\sigma})&\leq \frac{n}{\gamma \phi(n)k} + \sum_{i=0}^{k-1} C_i\\
& \leq \frac{n}{\gamma \phi(n)\log_4 n} + \mathcal{O}\left(\frac{n}{(\log n)^{3/2}}\right).
\end{align*}
This finishes the proof.
\end{proof}

The application of Theorem~\ref{thm:weakly-balanced} yields the following upper bound on the expected number of distinct fringe subtrees in random binary search trees:

\begin{example}\label{ex:bst-weaklybalanced}
For the binary search tree model from Example~\ref{ex:bst}, we find that $\sigma_{\text{bst}} \in \mathcal{L}_{\text{wbal}}(\phi_{\text{bst}}, \gamma_{\text{bst}})$ with $\phi_{\text{bst}}(n) = 1/2$ for every $n \geq 2$ and $\gamma_{\text{bst}} = 1/4$. Hence, by Theorem~\ref{thm:weakly-balanced}, we have
\begin{align*}
\mathbb{E}(F_{n,\sigma_{\text{bst}}})\leq \frac{16n}{\log n}+\mathcal{O}\left(\frac{n}{(\log n)^{3/2}}\right).
\end{align*}
The asymptotic growth coincides with the upper bound on $\mathbb{E}(F_{n,\sigma_{\text{bst}}})$ from Example~\ref{ex:upperbounded-bst} obtained by Theorem~\ref{thm:upper-bounded-theorem}, except for the leading constant. Recall the exact bound  \eqref{eq-upper-lower-BST-W24} for the binary search tree model
from \cite{Wa24}. So except for the leading constant, the upper bound from Theorem~\ref{thm:weakly-balanced} yields the correct asymptotic growth of $\mathbb{E}(F_{n,\sigma_{\text{bst}}})$ as well. Note that it is possible to choose $\phi_{\text{bst}}(n)=1-2\gamma$ for any $\gamma < 1/2$. For $\gamma=1/4$, we obtain the best leading constant in the upper bound.
\end{example}

\begin{example}\label{ex:dst-weaklybalanced}
For the binomial random tree model from Example~\ref{ex:dst}, we obtain the following result from Theorem~\ref{thm:weakly-balanced}. Let $S_{p}^n$ be the random variable taking values in the set $\{1,\dots, n-1\}$ according to the probability mass function $\sigma_{\text{bin},p}$ from \eqref{eq-dst}. Then $S_p^n = Z_p^n +1$, where $Z_p^n$ is a binomially distributed random variable with parameters $n-2$ and $p$. For the expected value $\mu$ of $S_p^n$, we thus obtain $\mu=\mathbb{E}(S_p^n)=p(n-2)+1$. Chernoff's bound implies
\begin{align*}
\mathbb{P}(|\mu - S_p^n| < \mu^{3/4}) \geq 1-2e^{-\frac{\sqrt{\mu}}{3}}.
\end{align*}
Moreover, with $i_1:=\mu-\mu^{3/4}$ and $i_2:=\mu+\mu^{3/4}$ we have
\begin{align*}
\mathbb{P}(|\mu - S_p^n| <\mu^{3/4}) = \sum_{i_1 < i < i_2} \sigma_{\text{bin},p}(i,n-i).
\end{align*}
Next, let $\gamma <\min\{p,1-p\} \le 1/2$.
There is an integer $N_{\sigma} $ (depending on $p$ and $\gamma$) such that 
\begin{align*}
\gamma n \leq \mu - \mu^{3/4}\leq \mu + \mu^{3/4} \leq (1-\gamma)n
\end{align*}
for all $n \geq N_{\sigma}$. All in all, we thus have
\begin{align*}
\sum_{\gamma n \leq i \leq n-\gamma n}\sigma_{\text{bin},p}(i,n-i)\geq \sum_{i_1 < i<i_2}\sigma_{\text{bin},p}(i,n-i) \geq 1-2e^{-\frac{\sqrt{p(n-2)+1}}{3}}
\end{align*}
for $n \geq N_{\sigma}$. Choose any constant $\varepsilon$ with  $0 < \varepsilon < 1$, and let
$\phi(n) = 1-\varepsilon$ (i.e., $\phi$ is a constant function). Then, if $n$ is large enough, we have
$$
\sum_{\gamma n \leq i \leq n-\gamma n}\sigma_{\text{bin},p}(i,n-i)\geq 1- \varepsilon,
$$
i.e., $\sigma_{\text{bin},p} \in \mathcal{L}_{\text{wbal}}(\phi, \gamma)$ for $N_{\sigma}$ large enough. From Theorem~\ref{thm:weakly-balanced}, we obtain
\begin{align*}
\mathbb{E}(F_{n,\sigma_{\text{bin},p}}) \leq c_{p}\cdot \frac{n}{\log n} + \mathcal{O}\left(\frac{n}{(\log n)^{3/2}}\right),
\end{align*}
where we can choose any constant
$c_{p}>2/p$ (if $p \leq 1/2$) respectively $c_{p}>2/(1-p)$ (if $p > 1/2$), as $\varepsilon>0$ is arbitrary. In particular, Theorem~\ref{thm:weakly-balanced} yields a more precise upper bound on $\mathbb{E}(F_{n,\sigma_{\text{bin},p}}) $ than Theorem~\ref{thm:upper-bounded-theorem} (see Example~\ref{ex:dst-upper-bounded}). In Example~\ref{ex:dst-information-theoretic}, we will show a corresponding lower bound of the form $\mathbb{E}(F_{n,\sigma_{\text{bin},p}}) \in \Omega(n/\log n)$. 
\end{example}

It remains to remark that for the uniform probability distribution (from Example~\ref{ex:uniform}), Theorem~\ref{thm:weakly-balanced} only yields a trivial upper bound on $\mathbb{E}(F_{n,\sigma_{\text{uni}}})$. On the one hand, we would need $\phi(n) \in \omega(1/\log n) $ in order to obtain a non-trivial upper bound from Theorem~\ref{thm:weakly-balanced}, but on the other hand, for every constant $0<\gamma < 1/2$, we find that if $\sigma_{\text{uni}} \in \mathcal{L}_{\text{wbal}}(\phi, \gamma)$, then $\phi(n) \in \mathcal{O}(1/\sqrt{n})$.
In the next section (Section~\ref{sec:strongly-balanced}), we present a class of leaf-centric binary tree sources which contains the uniform distribution $\sigma_{\text{uni}}$ from Example~\ref{ex:uniform}, and for which we will be able to derive a non-trivial asymptotic upper bound on $\mathbb{E}(F_{n,\sigma})$.

\subsection{Strongly-balanced sources}\label{sec:strongly-balanced}

In this section, we focus on another class of leaf-centric binary tree sources, which represents a refinement of the class of weakly-balanced leaf-centric tree sources from Section~\ref{sec:weaklybalanced}:

\begin{defi}[$\vartheta$-strongly-balanced sources] \label{def-sbal}
For a decreasing function $\vartheta: \mathbb{R} \to (0,1]$ and a constant $\gamma \in (0,\frac{1}{2})$, let $\mathcal{L}_{\text{sbal}}(\vartheta, \gamma)\subseteq \mathcal{L}$ denote the set of mappings $\sigma$ for which there is an integer $N_{\sigma}$ and a constant $c_{\sigma} \geq 1$ such that for every $n \geq N_{\sigma}$ and every integer $r$ with $c_{\sigma} \leq r \leq \lceil \gamma n\rceil$, the following inequality holds:
\begin{align*}
\sum_{r \leq i \leq n-r} \sigma(i,n-i)\geq \vartheta(r).
\end{align*}
\end{defi}

Let $\sigma \in \mathcal{L}_{\text{sbal}}(\vartheta, \gamma)$, and define $\phi\colon \mathbb{N} \to (0,1]$ by $\phi(n) = \vartheta( \lceil \gamma n\rceil)$. Then we have $\sigma \in \mathcal{L}_{\text{wbal}}(\phi, \gamma)$, i.e., every strongly-balanced leaf-centric tree source is also weakly-balanced. In particular, Theorem~\ref{thm:weakly-balanced} thus holds for strongly balanced tree sources as well.
However, we are able to prove a stronger asymptotic upper bound on the expected number $\mathbb{E}(F_{n,\sigma})$ of distinct fringe subtrees in a random tree $T_{n,\sigma}$ if $\sigma$ corresponds to a strongly-balanced tree source:

\begin{theorem}\label{thm:strongly-balanced}
Let $\vartheta: \mathbb{R} \to (0,1]$ be a decreasing function, let $\gamma \in (0,\frac{1}{2})$, and let $\sigma \in \mathcal{L}_{\text{sbal}}(\vartheta, \gamma)$. The number $F_{n,\sigma}$ of distinct fringe subtrees in the random tree $T_{n,\sigma}$ satisfies
\begin{align*}
\mathbb{E}(F_{n,\sigma})\leq  \frac{n}{\gamma \vartheta(\gamma \log_4 n)\log_4 n}(1+o(1)).
\end{align*}
\end{theorem}

In order to prove Theorem~\ref{thm:strongly-balanced}, we start again with an upper bound on the expected number $\mathbb{E}(Y_{n,k})$ of fringe subtrees of size larger than $k$ in a random tree $T_{n,\sigma}$, where $\sigma \in \mathcal{L}_{\text{sbal}}(\vartheta, \gamma)$. The following lemma and its proof are in fact quite similar to Lemma~\ref{lemma:weakly-balanced} and its proof:  

\begin{lemma}\label{lemma:strongly-balanced}
Let $\vartheta: \mathbb{R} \to (0,1]$ be a decreasing function, let $\gamma \in (0,\frac{1}{2})$, and let $\sigma \in \mathcal{L}_{\text{sbal}}(\vartheta, \gamma)$ with integer $N_{\sigma}$ and constant $c_{\sigma}$. Then for $k \geq \max\{N_{\sigma}, c_{\sigma}/\gamma\}$ and $n \geq k+1$, the (random) number $Y_{n,k}$ of fringe subtrees of size larger than $k$ in the random tree $T_{n, \sigma}$ satisfies
\begin{equation*}
\mathbb{E}(Y_{n,k}) \leq \frac{n}{\gamma \vartheta(\lceil \gamma (k+1)\rceil) k}-\frac{1}{\vartheta(\lceil \gamma (k+1)\rceil)}.
\end{equation*}
\end{lemma}

\begin{proof}
We prove the statement by induction on $n \geq k+1$. For the base case, let $n=k+1$. A binary tree $t$ of size $k+1$ has exactly one fringe subtree of size larger than $k$, thus
\begin{align*}
\mathbb{E}(Y_{k+1,k})=1\leq \frac{(\frac{1}{\gamma} -1)(k+1)}{\vartheta(\lceil \gamma (k +1)\rceil)k}\leq \frac{k+1}{\gamma\vartheta(\lceil \gamma (k+1)\rceil)k}-\frac{1}{\vartheta(\lceil \gamma (k+1)\rceil)},
\end{align*}
as $\frac{1}{\gamma}>2$ and $\vartheta(\lceil \gamma (k+1)\rceil)\leq 1$ by assumption. Let us now deal with the induction step. 
Take an integer $n > k+1$, so that 
\begin{equation} \label{eq-lemma:strongly-balanced-IH}
\mathbb{E}(Y_{i,k})\leq \frac{i}{\gamma \vartheta(\lceil \gamma (k+1)\rceil) k}-\frac{1}{\vartheta(\lceil \gamma (k+1)\rceil)}
\end{equation} 
for every integer $k+1 \leq i \leq n-1$ by the induction hypothesis. 
As $c_{\sigma} \leq \gamma k<\lceil \gamma (k+1)\rceil\leq \lceil \gamma n\rceil $ and $\sigma \in \mathcal{L}_{\text{sbal}}(\vartheta, \gamma)$, we have 
\begin{equation} \label{ineq-D_n-sbal}
D_n(x) \leq 1-\vartheta(\lceil \gamma (k+1)\rceil)
\end{equation}
for every $0 \leq x \leq \gamma k$ (take $r = \lceil \gamma (k+1)\rceil$ in Definition~\ref{def-sbal}).  
By Lemma~\ref{lemma:recurrence-expectation}, we have
\begin{align*}
\mathbb{E}(Y_{n,k})=1+\sum_{1 \leq i \leq n/2}\sigma^*(i,n-i)(\mathbb{E}(Y_{i,k})+\mathbb{E}(Y_{n-i,k})).
\end{align*}
First, note that for all $1 \leq i \leq n/2$ we have
\begin{align*}
\mathbb{E}(Y_{i,k})\leq \max\left\{\frac{i}{\gamma \vartheta(\lceil \gamma (k+1)\rceil)k}-\frac{1}{\vartheta(\lceil \gamma (k+1)\rceil)},0\right\}.
\end{align*}
This follows from \eqref{eq-lemma:strongly-balanced-IH} for $i > k$, and from the fact that $\mathbb{E}(Y_{i,k})=0$ for $i \leq k$. 
Furthermore, for all $1 \leq i \leq n/2$ we have
\begin{align*}
\mathbb{E}(Y_{n-i,k})\leq \frac{n-i}{\gamma \vartheta(\lceil \gamma (k+1)\rceil) k}-\frac{1}{\vartheta(\lceil \gamma (k+1)\rceil)}.
\end{align*}
This follows again from \eqref{eq-lemma:strongly-balanced-IH} if $n-i >k$. For $n-i \leq k$ note that
\begin{align*}
\mathbb{E}(Y_{n-i,k}) = 0 &< \frac{n-2\gamma k}{2\gamma \vartheta(\lceil \gamma (k+1)\rceil)k}=\frac{n}{2\gamma\vartheta(\lceil \gamma (k+1)\rceil)k}-\frac{1}{\vartheta(\lceil \gamma (k+1)\rceil)} \\
&\le \frac{n-i}{\gamma \vartheta(\lceil \gamma (k+1)\rceil)k}-\frac{1}{\vartheta(\lceil \gamma (k+1)\rceil)},
\end{align*}
since $i \le n/2$ and $\gamma < 1/2$ by assumption.  Thus, we obtain
\begin{align*}
\mathbb{E}(Y_{n,k}) & \leq 1+\sum_{1 \leq i\leq n/2}\sigma^*(i,n-i) \bigg(
\max\left\{\frac{i}{\gamma \vartheta(\lceil \gamma (k+1)\rceil)k}-\frac{1}{\vartheta(\lceil \gamma (k+1)\rceil)},0\right\}
\\
&\qquad + \frac{n-i}{\gamma \vartheta(\lceil \gamma (k+1)\rceil)k}-\frac{1}{\vartheta(\lceil \gamma (k+1)\rceil)}
\bigg)
\\
&= 1+\frac{n}{\gamma \vartheta(\lceil \gamma (k+1)\rceil)k}-\frac{1}{\vartheta(\lceil \gamma (k+1)\rceil)}\\
&\qquad -\sum_{1 \leq i\leq n/2}\sigma^*(i,n-i)\min\left\{\frac{i}{\gamma \vartheta(\lceil \gamma (k+1)\rceil)k}, \frac{1}{\vartheta(\lceil \gamma (k+1)\rceil)}\right\} \\
\overset{\text{Lem.~\ref{lemma-integral}}}&{=} 1+\frac{n}{\gamma \vartheta(\lceil \gamma (k+1)\rceil)k}-\frac{2}{\vartheta(\lceil \gamma (k+1)\rceil)} +\int_0^{\gamma k}\frac{D_n(x)}{\gamma \vartheta(\lceil \gamma (k+1)\rceil)k} \, \mathrm{d}x  \\
\overset{\text{\eqref{ineq-D_n-sbal}}}&{\leq} 1+\frac{n}{\gamma \vartheta(\lceil \gamma (k+1)\rceil) k}-\frac{2}{\vartheta(\lceil \gamma (k+1)\rceil)} +\int_0^{\gamma k}\frac{1-\vartheta(\lceil \gamma (k+1)\rceil)}{\gamma \vartheta(\lceil \gamma (k+1)\rceil)k}  \, \mathrm{d}x \\
& = \frac{n}{\gamma \vartheta(\lceil \gamma (k+1)\rceil )k}-\frac{1}{\vartheta(\lceil \gamma (k+1)\rceil)},
\end{align*}
which concludes the proof.
\end{proof}

We are now able to prove Theorem~\ref{thm:strongly-balanced}.

\begin{proof}[Proof of Theorem~\ref{thm:strongly-balanced}]
Let  $m=\max\{N_{\sigma},c_{\sigma}/\gamma\}$.
By Lemma~\ref{lemma:strongly-balanced}, the expected number $\mathbb{E}(Y_{n,k})$ of fringe subtrees of size larger than $k$ in $T_{n,\sigma}$ satisfies
\begin{align*}
\mathbb{E}(Y_{n,k})\leq \frac{n}{\gamma \vartheta(\lceil \gamma (k+1)\rceil)k}-\frac{1}{\vartheta(\lceil \gamma (k+1)\rceil)}
\end{align*}
for all $n > k \geq m$. Set $k := \lfloor \log_4 n-1/\gamma \rfloor-1$ (this implies $\vartheta(\lceil \gamma (k+1)\rceil) \ge \vartheta(\gamma \log_4 n)$),
and let $n$ be sufficiently large, so that $k \geq m$. By the cut-point argument from Section~\ref{sec:preliminaries} (Lemma~\ref{lemma:upper-bound-Fn}), we find that
\begin{align*}
\mathbb{E}(F_{n,\sigma})\leq \mathbb{E}(Y_{n,k}) + \sum_{i=0}^{k-1}C_i.
\end{align*}
With the asymptotic growth of the Catalan numbers \eqref{eq:Catalan-growth}, and with Lemma~\ref{lemma:strongly-balanced}, we find
\begin{align*}
\mathbb{E}(F_{n,\sigma}) &\leq \frac{n}{\gamma \vartheta(\lceil \gamma (k+1)\rceil)k}+\mathcal{O}\left(\frac{4^k}{k^{3/2}}\right) \\\
&\leq  \frac{n}{\gamma \vartheta(\gamma \log_4 n)\log_4 n}(1+o(1)).
\end{align*}
This finishes the proof.
\end{proof}

With Theorem~\ref{thm:strongly-balanced}, we obtain a non-trivial asymptotic upper bound on the expected number of distinct fringe subtrees in a uniformly random binary tree of size $n$:

\begin{example}\label{ex:uniform-strongly-balanced}
Consider the uniform distribution corresponding to $\sigma_{\text{uni}}$ from Example~\ref{ex:uniform}.
Let $0 <\gamma < 1/2$, and let $\varepsilon > 0$. From the asymptotic bound of the Catalan numbers \eqref{eq:Catalan-growth}, we find that there is an integer $N_\varepsilon$ (depending on $\varepsilon$) such that
\begin{align*}
\frac{(1-\varepsilon)4^n}{\sqrt{\pi}(n+1)^{3/2}} \leq C_n \leq \frac{(1+\varepsilon)4^n}{\sqrt{\pi}(n+1)^{3/2}}
\end{align*}
for all $n \geq N_\varepsilon$. Let $N_{\varepsilon} \leq r \leq \lceil \gamma n \rceil $, then
\begin{align*}
\sum_{r \leq i \leq n-r}\sigma_{\text{uni}}(i,n-i) &= \sum_{r \leq i \leq n-r}\frac{C_{i-1}C_{n-i-1}}{C_{n-1}}\\
&\geq \frac{(1-\varepsilon)^2}{4(1+\varepsilon)\sqrt{\pi}}\sum_{r \leq i \leq n-r}\frac{n^{3/2}}{i^{3/2}(n-i)^{3/2}}\\
&\geq \frac{(1-\varepsilon)^2n^{3/2}}{2(1+\varepsilon)\sqrt{\pi}}\sum_{r \leq i < \frac{n}{2}}i^{-3/2}(n-i)^{-3/2}\\
&\geq \frac{(1-\varepsilon)^2n^{3/2}}{2(1+\varepsilon)\sqrt{\pi}}\int_r^{\frac{n}{2}}x^{-3/2}(n-x)^{-3/2}\mathrm{d}x\\
&=\frac{(1-\varepsilon)^2}{(1+\varepsilon)\sqrt{\pi}} \frac{n-2r}{\sqrt{n(n-r)r}}.
\end{align*}
If $N_{\varepsilon}$ is chosen large enough, then we also have
$$\frac{n-2r}{\sqrt{n(n-r)}} \leq 1 - 2\gamma$$
for all $r,n$ with $n \geq N_{\varepsilon}$ and $r \leq \lceil n\gamma \rceil $. So set 
$\delta:=\frac{(1+\varepsilon)}{(1-\varepsilon)^2}$, and define 
\begin{align*}
\vartheta_{uni}(r) :=\frac{(1-2\gamma)}{\delta\sqrt{\pi r}}.
\end{align*}
Then $\sigma_{\text{uni}} \in \mathcal{L}_{\text{sbal}}(\vartheta_{uni}, \gamma)$. 
By Theorem~\ref{thm:strongly-balanced}, we thus obtain
\begin{align*}
\mathbb{E}(F_{n,\sigma_{\text{uni}}}) \leq \frac{\delta \sqrt{2\pi}}{\sqrt{\gamma} (1-2\gamma)}\cdot \frac{n}{\sqrt{\log n}} (1+o(1)).
\end{align*}
As $0 < \gamma < 1/2$ is arbitrary, we can choose the optimal value $\gamma = 1/6$ to obtain 
\begin{align*}
\mathbb{E}(F_{n,\sigma_{\text{uni}}}) \leq 3\delta\sqrt{3\pi}\cdot \frac{n}{\sqrt{\log n}}(1+o(1)),
\end{align*}
where $\delta>1$, as $\varepsilon>0$.
In \cite{FlajoletSS90,RalaivaosaonaW15,SeelbachW22}, it is shown that 
$$\mathbb{E}(F_{n,\sigma_{\text{uni}}})=\frac{2\sqrt{2}}{\sqrt{\pi}}\cdot \frac{n}{\sqrt{\log n}}(1+o(1)).$$ Thus, except for the 
leading constant, the upper bound from Theorem~\ref{thm:strongly-balanced} yields the correct asymptotic growth of $\mathbb{E}(F_{n,\sigma_{\text{uni}}})$.
\end{example}

\begin{example}\label{ex:bst-stronglybalanced}
For the binary search tree model $\sigma_{\text{bst}}$ from Example~\ref{ex:bst}, we also obtain an upper bound on $\mathbb{E}(F_{n,\sigma_{\text{bst}}})$ from Theorem~\ref{thm:strongly-balanced}. Let $\gamma_{\text{bst}}=1/4$, let $n \geq 2$, and let $1 \leq r \leq \lceil \gamma n\rceil $. Then
\begin{align*}
\sum_{r \leq i \leq n-r}\sigma_{\text{bst}}(i,n-i)= \frac{n-2r+1}{n-1}\geq \frac{1}{2}.
\end{align*}
Set $\vartheta_{\text{bst}}(r):=1/2$.
By Theorem~\ref{thm:strongly-balanced}, we find
\begin{align*}
\mathbb{E}(F_{n,\sigma_{\text{uni}}}) \leq \frac{16n}{\log n} (1+o(1)).
\end{align*}
This coincides with the asymptotic upper bound on $\mathbb{E}(F_{n,\sigma_{\text{uni}}})$ from Example~\ref{ex:bst-weaklybalanced} as well as Example~\ref{ex:upperbounded-bst} (except for the leading constant).
\end{example}

\section{Lower Bounds}

So far, we have only considered upper bounds on the expected number of distinct fringe subtrees in random trees $T_{n,\sigma}$. In this section, we focus on lower bounds instead. We present some classes of leaf-centric binary tree sources for which we will be able to show asymptotic lower bounds on the number $\mathbb{E}(F_{n,\sigma})$ of distinct fringe subtrees in the corresponding random trees $T_{n,\sigma}$.

\subsection{Upper-bounded sources}\label{sec:information-theoretic}

First, we consider a specific subclass of the class of upper-bounded leaf-centric tree sources from Definition \ref{def:upper-bounded}.
Let $\rho \in (0,1)$ be a constant. With $\mathcal{L}_{\text{up}}(\rho)\subseteq \mathcal{L}$, we denote the set of mappings $\sigma \in \mathcal{L}$ that are eventually bounded above by the constant $\rho$: that is, a mapping $\sigma \in \mathcal{L}$ belongs to $\mathcal{L}_{\text{up}}(\rho)$ if there is an integer $N_{\sigma}$ such that 
\begin{align*}
\sigma(i,n-i)\leq \rho
\end{align*}
for all $n \geq N_{\sigma}$ and all integers $1 \leq i \leq n-1$. 
In the following theorem, we obtain a lower bound on $\mathbb{E}(F_{n,\sigma})$ for $\sigma \in \mathcal{L}_{\text{up}}(\rho)$.

\begin{theorem}\label{thm:lower-bound-information-theoretic}
Let $\rho \in (0,1)$, and let $\sigma \in \mathcal{L}_{\text{up}}(\rho)$ with integer $N_{\sigma}$. Then the number $F_{n,\sigma}$ of distinct fringe subtrees in the random tree $T_{n,\sigma}$ satisfies
\begin{align*}
\mathbb{E}(F_{n,\sigma}) \geq \log \left(\frac{1}{\rho}\right)\frac{n}{2(N_{\sigma}-1)\log n}\left(1+o(1)\right).
\end{align*}
\end{theorem}
In order to prove Theorem~\ref{thm:lower-bound-information-theoretic}, we make use of an information-theoretic argument. Let $H(T_{n,\sigma})$ denote the Shannon entropy of the random variable $T_{n,\sigma}$, i.e.,
\begin{align*}
H(T_{n,\sigma})=\sum_{t \in \mathcal{T}_n}P_{\sigma}(t)\log\left(1/P_{\sigma}(t)\right).
\end{align*}
We obtain the following general result.

\begin{theorem}\label{thm:lower-bound-with-entropy}
Let $\sigma \in \mathcal{L}$. Then the number $F_{n,\sigma}$ of distinct fringe subtrees in the random tree $T_{n,\sigma}$ satisfies
\begin{align*}
\mathbb{E}(F_{n,\sigma})\geq \frac{H(T_{n,\sigma})}{2\log n}\left(1+o(1)\right).
\end{align*}
\end{theorem}

\begin{proof}
Let $F(t)$ denote the number of distinct fringe subtrees occurring in the binary tree $t \in \mathcal{T}$. 
Recall that $t \in \mathcal{T}_n$ has exactly $n$ leaves and $n-1$ internal nodes. This implies that
$F(t) \leq n$ for $t \in \mathcal{T}_n$.
We first show that we need at most $2F(t)(\lfloor \log n \rfloor +1)$ many bits in order to encode $t$. Recall that $F(t)$ equals the size of the minimal DAG of $t$, i.e., the number of nodes in the directed acyclic graph obtained from $t$ by merging identical fringe subtrees of $t$. As we can uniquely retrieve $t$ from its minimal DAG (see e.g.~\cite{LohreyMR17}), it suffices to show that we can encode the minimal DAG of $t$ with at most $2F(t)(\lfloor\log n \rfloor +1)$ many bits. Without loss of generality, assume that the nodes of the minimal DAG of $t$ are enumerated as $1,2, \dots, F(t)$, where $F(t)$ is the unique leaf node of the minimal DAG of $t$. For $1 \leq i \leq F(t)-1$, let $\ell_i$ (resp.~$r_i$) denote the left (resp.~right) child of node $k$.  We encode each number $1, \dots, F(t)$ by a bit string of length exactly $\lfloor \log n\rfloor +1$. The minimal DAG of $t$ can be uniquely encoded as the bit string $\ell_1r_1\cdots \ell_{F(t)-1}r_{F(t)-1}$, which has length $2(F(t)-1)(\lfloor \log n \rfloor+1)$.
Let $\sigma \in \mathcal{L}$.
Shannon's coding theorem implies
\begin{align*}
H(T_{n,\sigma})\leq 2(\lfloor \log n \rfloor+1)\sum_{t \in \mathcal{T}_n}P_{\sigma}(t)F(t)=2(\lfloor \log n \rfloor+1)\mathbb{E}(F_{n,\sigma}).
\end{align*}
This finishes the proof.
\end{proof}

In order to show Theorem~\ref{thm:lower-bound-information-theoretic}, we make use of Theorem~\ref{thm:lower-bound-with-entropy} together with a lower bound on the entropy $H(T_{n,\sigma})$ for $\sigma \in \mathcal{L}_{\text{up}}(\rho)$. We will show the following lemma.

\begin{lemma}\label{lemma:entropy-lower-bound}
Let $\rho \in (0,1)$, and let $\sigma \in \mathcal{L}_{\text{up}}(\rho)$ with integer $N_{\sigma} \ge 2$. Then
\begin{align*}
H(T_{n,\sigma})\geq \log \left(\frac{1}{\rho}\right)\left(\frac{n}{N_{\sigma}-1}-1\right)
\end{align*}
for every $n \geq N_{\sigma}$. 
\end{lemma}

In order to prove Lemma~\ref{lemma:entropy-lower-bound}, we need the following result:

\begin{lemma}\label{lemma:lower-bound-leafsize}  Let $k \geq 1$.
Every binary tree of size $n$ contains at least $\frac{n}{k} - 1$ fringe subtrees of size larger than $k$. 
\end{lemma}
\begin{proof}
Let $t$ be a binary tree of size $n$, and let $s_1,s_2,\ldots,s_r$ be the maximal fringe subtrees of size at most $k$ (i.e., the fringe subtrees that have at most $k$ leaves and are not contained in any other fringe subtree with that property). Every leaf of $t$ is contained in one of these: since a leaf is itself a fringe subtree of size at most $k$ (namely size $1$), it must be contained in a maximal fringe subtree with that property. Moreover, $s_1,s_2,\ldots,s_r$ must be disjoint, since for any two fringe subtrees of the same tree, either one is contained in the other, or they are disjoint. Hence we have $n = |t| = |s_1| + |s_2| + \cdots + |s_r| \leq kr$, which implies $r \geq \frac{n}{k}$. Lastly, note that the roots of $s_1,s_2,\ldots,s_r$ and all internal nodes of $t$ that are not contained in these $r$ fringe subtrees form a binary tree of size $r$. This tree has $r-1$ internal nodes, and all of them are roots of fringe subtrees of $t$ whose size is at least $k+1$.
\end{proof}

In particular, we obtain the following corollary from Lemma~\ref{lemma:lower-bound-leafsize}:

\begin{cor}\label{cor:lower-bound-eynk}
Let $\sigma \in \mathcal{L}$, and let $n>k$. The random number $Y_{n,k}$ of fringe subtrees of size larger than $k$ in the random tree $T_{n,\sigma}$ satisfies
\begin{align*}
\mathbb{E}(Y_{n,k}) \geq \frac{n}{k}-1.
\end{align*}
\end{cor}

We are now able to prove Lemma~\ref{lemma:entropy-lower-bound}:
\begin{proof}[Proof of Lemma~\ref{lemma:entropy-lower-bound}]
Lemma~\ref{lemma:entropy-lower-bound} follows from identity (4) in \cite{KiefferYS09}: Define 
\begin{align*}
h_k(\sigma)=\sum_{i=1}^{k-1}\sigma(i,k-i)\log\left(\frac{1}{\sigma(i,k-i)}\right).
\end{align*}
Thus, $h_k(\sigma)$ is the Shannon entropy of the random variable corresponding to the probability mass function $\sigma: \{(i,k-i) \mid 1 \leq i \leq k-1\} \to [0,1]$. As $\sigma(i,k-i)\leq \rho$ for $k \geq N_{\sigma}$, we find
\begin{align*}
h_k(\sigma)\geq \log\left(1/\rho\right)\sum_{i=1}^{k-1}\sigma(i,k-i)=\log\left(1/\rho\right)
\end{align*}
for every $k \geq N_{\sigma}$. Identity (4) in \cite{KiefferYS09} states that
\begin{align*}
H(T_{n,\sigma}) = \sum_{k=2}^n\left(\mathbb{E}(Y_{n,k-1})-\mathbb{E}(Y_{n,k})\right)h_k(\sigma),
\end{align*}
where $Y_{n,k}$ again denotes the random number of fringe subtrees of size larger than $k$ in the random tree $T_{n,\sigma}$.
For every $n \geq N_{\sigma} \ge 2$, we obtain 
\begin{align*}
H(T_{n,\sigma}) &\geq  \sum_{k=N_{\sigma}}^n\left(\mathbb{E}(Y_{n,k-1})-\mathbb{E}(Y_{n,k})\right)h_k(\sigma)\\
&\geq \log \left(\frac{1}{\rho}\right)\sum_{k=N_{\sigma}}^n\left(\mathbb{E}(Y_{n,k-1})-\mathbb{E}(Y_{n,k})\right)\\
&= \log \left(\frac{1}{\rho}\right)\left(\mathbb{E}(Y_{n,N_{\sigma}-1})-\mathbb{E}(Y_{n,n})\right)\\
&=\log \left(\frac{1}{\rho}\right)\mathbb{E}(Y_{n,N_{\sigma}-1}).
\end{align*}
By Corollary \ref{cor:lower-bound-eynk}, this implies
\begin{align*}
H(T_{n,\sigma}) &\geq \log \left(\frac{1}{\rho}\right)\left(\frac{n}{N_{\sigma}-1}-1\right).
\end{align*}
This proves the statement.
\end{proof}

Theorem~\ref{thm:lower-bound-information-theoretic} now follows immediately from Theorem~\ref{thm:lower-bound-with-entropy} and Lemma~\ref{lemma:entropy-lower-bound}.

For all the random tree models from Examples~\ref{ex:bst}--\ref{ex:beta} one can obtain with Theorem~\ref{thm:lower-bound-information-theoretic} lower bounds of $\Omega(n / \log n)$ for the average number of distinct fringe subtrees. In some cases, one obtains slightly better multiplicative
constants by directly using Theorem~\ref{thm:lower-bound-with-entropy}.

\begin{example}
Let us apply Theorem~\ref{thm:lower-bound-with-entropy} to the binary search tree model from Example~\ref{ex:bst}. In \cite{KiefferYS09}, it is shown that 
\begin{align*}
H(T_{n,\sigma_{\text{bst}}})\sim 1.7363771368 \, n \, (1+o(1)).
\end{align*}
Hence, we obtain from Theorem~\ref{thm:lower-bound-with-entropy} that
\begin{align*}
\mathbb{E}(F_{n,\sigma_{\text{bst}}})\geq \frac{0.8681 \, n}{\log n}(1+o(1)).
\end{align*}
In Example~\ref{ex:upperbounded-bst}, we have shown the upper bound
\begin{align*}
\mathbb{E}(F_{n,\sigma_{\text{bst}}})\leq \frac{8n}{\log n}(1+o(1))
\end{align*}
(see also Examples \ref{ex:bst-weaklybalanced} and  \ref{ex:bst-stronglybalanced}). Altogether, we find that $\mathbb{E}(F_{n,\sigma}) \in \Theta(n/\log n)$ (as shown in \cite{Devroye98, SeelbachW22} and improved in \cite{Wa24} with respect to the exact multiplicative constant).
\end{example}

\begin{example}\label{ex:dst-information-theoretic}
For the binomial random tree model (Example~\ref{ex:dst}), a simple induction over $n$ (using $\binom{n-2}{i-1} = \binom{n-3}{i-1} + \binom{n-3}{i-2}$ for $2 \le i \le n-2$ in the induction step)
 shows that for $n \geq 3$, we have
\begin{align*}
\sigma_{\text{bin},p}(i,n-i) \leq \max\{p,1-p\}
\end{align*}
for all $1 \leq i \leq n-1$. With $N_{\sigma}=3$ and $\rho=\max\{p,1-p\}$, we thus obtain from Theorem~\ref{thm:lower-bound-information-theoretic} that
\begin{align}\label{eq:lower-bound-dst1}
\mathbb{E}(F_{n,\sigma_{\text{bin},p}})\geq \log\left(\frac{1}{p}\right)\frac{n}{4\log n}(1+o(1))
\end{align}
if $p > 1/2$ and
\begin{align}\label{eq:lower-bound-dst2}
\mathbb{E}(F_{n,\sigma_{\text{bin},p}})\geq \log\left(\frac{1}{1-p}\right)\frac{n}{4\log n}(1+o(1))
\end{align}
if $p \leq 1/2$. 
If we consider the proof of Lemma~\ref{lemma:entropy-lower-bound} again, we can slightly improve the leading constant in the lower bounds \eqref{eq:lower-bound-dst1} and \eqref{eq:lower-bound-dst2}.  With identity (4) in \cite{KiefferYS09}, $h_2(\sigma)=0$, and Corollary \ref{cor:lower-bound-eynk}, we get
\begin{align*}
H(T_{n,\sigma}) &= \sum_{k=2}^n\left(\mathbb{E}(Y_{n,k-1})-\mathbb{E}(Y_{n,k})\right)h_k(\sigma)\\
&\geq \min_{3 \leq i \leq n}h_i(\sigma)\cdot \sum_{k=3}^n\left(\mathbb{E}(Y_{n,k-1})-\mathbb{E}(Y_{n,k})\right)\\
&=\min_{3 \leq i \leq n}h_i(\sigma)\cdot \mathbb{E}(Y_{n,2})\\
&\ge \min_{3 \leq i \leq n}h_i(\sigma) \cdot \left(\frac{n}{2}-1\right) .
\end{align*}
Together with Theorem~\ref{thm:lower-bound-with-entropy}, we thus obtain
\begin{align*}
\mathbb{E}(F_{n,\sigma})\geq \min_{3 \leq i \leq n}h_i(\sigma) \cdot\frac{n}{4\log n}(1+o(1)).
\end{align*}
For the binomial random tree model we have
$h_k(\sigma_{\text{bin},p}) \leq h_{k+1}(\sigma_{\text{bin},p})$: if $\mathrm{Ber}(p)$ is a Bernoulli random variable that takes the value $1$ (resp., $0$)
with probability $p$ (resp., $1-p$), then the binomial random variable $\mathrm{Bin}(k,p)$ is the sum of $k$ independent copies of $\mathrm{Ber}(p)$.
Hence, we have $\mathrm{Bin}(k+1,p) = \mathrm{Bin}(k,p) + \mathrm{Ber}(p)$. Moreover, for independent random variables $X$ and $Y$ the Shannon entropy satisfies $H(X+Y) \geq H(X)$; see e.g. \cite{Madiman08}.
Hence, we get
\begin{align*}
\min_{3 \leq i \leq n}h_i(\sigma_{\text{bin},p})=h_3(\sigma_{\text{bin},p})=p\log\left(\frac{1}{p}\right)+(1-p)\log\left(\frac{1}{1-p}\right).
\end{align*}
This yields
\begin{align*}
\mathbb{E}(F_{n,\sigma_{\text{bin},p}})\geq \left(p\log\left(\frac{1}{p}\right)+(1-p)\log\left(\frac{1}{1-p}\right)\right) \cdot\frac{n}{4\log n}(1+o(1)).
\end{align*}
 Together with the upper bound (Example~\ref{ex:dst-weaklybalanced}), we get $\mathbb{E}(F_{n,\sigma_{\text{bin},p}}) \in \Theta(n/\log n)$.
\end{example}

\begin{example}
For the uniform distribution from Example~\ref{ex:uniform}, we have (see \cite{KiefferYS09})
\begin{align*}
H(T_{n,\sigma_{\text{uni}}})\sim 2n(1+o(1)).
\end{align*}
Theorem~\ref{thm:lower-bound-with-entropy} thus yields a lower bound 
\begin{align*}
\mathbb{E}(F_{n,\sigma_{\text{uni}}})\geq \frac{n}{\log n}(1+o(1)).
\end{align*}
However, in \cite{FlajoletSS90, RalaivaosaonaW15}, it is shown that 
\begin{align*}
\mathbb{E}(F_{n,\sigma_{\text{uni}}})=\frac{2\sqrt{2}}{\sqrt{\pi}}\cdot \frac{n}{\sqrt{\log n}}\left(1+o(1)\right).
\end{align*}
In the next subsection, we will prove a general result that implies a lower bound of the form $\mathbb{E}(F_{n,\sigma_{\text{uni}}})\in \Omega\left(n/\sqrt{\log n}\right)$ (see Example~\ref{ex:uniform-unbalanced}).
\end{example}

\subsection{Unbalanced sources}

For upper-bounded sources, Theorem~\ref{thm:lower-bound-information-theoretic} can only yield lower bounds of the form $\Omega(n/\log n)$ for $\mathbb{E}(F_{n,\sigma})$.
In this subsection, we present another class of leaf-centric binary tree sources, for which we will be able to derive stronger lower bounds on $\mathbb{E}(F_{n,\sigma})$, such as the lower bound  $\Omega(n/\sqrt{\log n})$ for the uniform distribution.
We start with the following definition:

\begin{defi}[$\xi$-unbalanced sources]\label{def:unbalanced}
Let $\xi \colon \mathbb{R} \to (0,1]$ be a decreasing function, and let $\gamma \in (0,\frac{1}{2})$. With $\mathcal{L}_{unbal}(\xi, \gamma) \subseteq \mathcal{L}$, we denote the set of mappings $\sigma$ for which there are integers $N'_{\sigma}$ and $c_{\sigma}$ such that
\begin{align*}
\sum_{r \leq i \leq n-r}\sigma(i,n-i)\leq \xi(r)
\end{align*}
for every $n \geq N'_{\sigma}$ and every $c_{\sigma} \leq r \leq \lceil \gamma n \rceil$.
\end{defi}

Moreover, let $\rho \in (0,1)$ be a constant. As in Section~\ref{sec:information-theoretic}, let again $\mathcal{L}_{\text{up}}(\rho) \subseteq \mathcal{L}$ denote the set of mappings $\sigma \in \mathcal{L}$ that are upper-bounded with respect to the constant $\rho$. In other words, a mapping $\sigma \in \mathcal{L}$ is contained in $\mathcal{L}_{\text{up}}(\rho)$ if there is an integer $N_{\sigma}$ such that
\begin{align*}
\sigma(i,n-i) \leq \rho
\end{align*}
for all $n \geq N_{\sigma}$ and $1 \leq i \leq n-1$.

Our main result in this subsection is the following theorem:

\begin{theorem}\label{thm:lower-bound-unbalanced}
Let $\xi$ be a decreasing function, let $\gamma \in (0,\frac{1}{2})$ and $\rho \in (0,1)$. Moreover, let $\sigma \in \mathcal{L}_{\text{up}}(\rho)$ with integer $N_{\sigma}$, and let $\sigma \in \mathcal{L}_{unbal}(\xi, \gamma)$ with integer $N'_{\sigma}$ and constant $c_{\sigma}$. If 
\begin{itemize}
\item[(a)] $\xi(x) \in \omega(1/x)$, 
\item[(b)] there is a constant $c_{\xi}>0$ such that $1/c_{\xi}+\gamma \leq 1$ and
\begin{align*}
\sum_{i=c_{\sigma}}^k\xi(i) \leq c_{\xi}(k+1)\xi(k+1)
\end{align*}
for every $k\geq c_{\sigma}$, and
\item[(c)] there is an integer $\ell_{\sigma}\geq 2$ such that the number $Y_{n,k}$ of fringe subtrees of size larger than $k$ in the random tree $T_{n,\sigma}$ satisfies $$\mathbb{E}(Y_{n,(\log n)^{\ell_{\sigma}}})\in o(n/\log n),$$
\end{itemize}
then the number $F_{n,\sigma}$ of distinct fringe subtrees in the random tree $T_{n,\sigma}$ satisfies
\begin{align*}
\mathbb{E}(F_{n,\sigma})\geq \frac{\gamma n\log(1/\rho)}{c_{\xi}N_{\sigma}\xi\big(\frac{N_{\sigma}\log n}{\log(1/\rho)}\big)\log n}\left(1+o(1)\right).
\end{align*}
\end{theorem}

Note that every leaf-centric binary tree source is unbalanced with respect to the constant function $\xi$ with $\xi(x)=1$ for every $x \in \mathbb{R}$. This yields the following corollary of Theorem~\ref{thm:lower-bound-unbalanced}, which is a weaker version of Theorem~\ref{thm:lower-bound-information-theoretic}:

\begin{cor}\label{cor:unbalanced}
Let $\rho \in (0,1)$, and let $\sigma \in \mathcal{L}_{\text{up}}(\rho)$ with integer $N_{\sigma}$. If there is an integer $\ell_{\sigma}\geq 2$ such that the number $Y_{n,k}$ of fringe subtrees of size larger than $k$ in the random tree $T_{n,\sigma}$ satisfies
\begin{align*}
\mathbb{E}(Y_{n,(\log n)^{\ell_{\sigma}}}) \in o(n/\log n),
\end{align*}
then the number $F_{n,\sigma}$ of distinct fringe subtrees in the random tree $T_{n,\sigma}$ satisfies
\begin{align*}
\mathbb{E}(F_{n,\sigma})\geq \frac{\gamma n\log(1/\rho)}{2N_{\sigma}\log n}(1+o(1)),
\end{align*}
for every $\gamma \in (0, \frac{1}{2})$.
\end{cor}
\begin{proof}
Let $\sigma \in \mathcal{L}$ satisfy the requirements of Corollary \ref{cor:unbalanced}. We show that $\sigma$ then satisfies all further requirements of Theorem~\ref{thm:lower-bound-unbalanced} as well.

Clearly, $\sigma$ is $\xi_1$-unbalanced for every constant $\gamma \in (0,1/2)$, where $\xi_1$ is the constant function defined by $\xi_1(x)=1$ for every $x \in \mathbb{R}$. Hence, we have $\sigma \in \mathcal{L}_{unbal}(\xi_1,\gamma)$ with $N_{\sigma}'=1$ and $c_{\sigma}=1$. We now find that condition (a) of Theorem~\ref{thm:lower-bound-unbalanced} is clearly satisfied, as is condition (b), where the constant $c_{\xi}$ can be chosen for example as $c_{\xi}=2$. Finally, condition (c) of Theorem~\ref{thm:lower-bound-unbalanced} is satisfied by assumption. Thus, we obtain the lower bound
\begin{align*}
\mathbb{E}(F_{n,\sigma})\geq \frac{\gamma n\log(1/\rho)}{2N_{\sigma}\log n}(1+o(1))
\end{align*}
from Theorem~\ref{thm:lower-bound-unbalanced}, where $\gamma \in (0,1/2)$ is arbitrary.
\end{proof}

Note that Corollary \ref{cor:unbalanced} resembles Theorem~\ref{thm:lower-bound-information-theoretic} in many ways, except that we have the additional requirement (c) from Theorem~\ref{thm:lower-bound-unbalanced} and the additional constant $\gamma \in (0,1/2)$ in the lower bound. The technique to prove Theorem~\ref{thm:lower-bound-unbalanced} however will be quite different from the technique used for Theorem~\ref{thm:lower-bound-information-theoretic}.

In order to prove Theorem~\ref{thm:lower-bound-unbalanced}, we make use of a refinement of the cut-point technique from Section~\ref{sec:preliminaries} (Lemma~\ref{lemma:upper-bound-Fn}), as was applied to a similar problem in \cite{RalaivaosaonaW15}. More precisely,
we refine the cut-point technique with an inclusion-exclusion-principle-like argument in order to obtain a lower bound on the expected number of distinct fringe subtrees.
We first need a lower bound on the expected number $\mathbb{E}(Y_{n,k})$ of fringe subtrees of size larger than $k$ in a random tree $T_{n,\sigma}$, where $\sigma \in \mathcal{L}_{unbal}(\xi, \gamma)$: 

\begin{lemma}\label{lemma:unbalanced}
Let $\xi \colon \mathbb{R} \to (0,1]$ be a decreasing function, let $\gamma \in (0,\frac{1}{2})$, and let $\sigma \in \mathcal{L}_{unbal}(\xi, \gamma)$ with integers $N'_{\sigma}$ and $c_{\sigma}$. If
\begin{itemize}
\item[(a)] $\xi(x) \in \omega(1/x)$, and
\item[(b)] there is a constant $c_{\xi}>0$ such that $1/c_{\xi}+\gamma \leq 1$ and
\begin{align*}
\sum_{i=c_{\sigma}}^k\xi(i) \leq c_{\xi}(k+1)\xi(k+1)
\end{align*}
for every $k \geq c_{\sigma}$, 
\end{itemize}
then there is an integer $m$ such that for $k \geq m$ and $n \geq k+1$, the random number $Y_{n,k}$ of fringe subtrees of size larger than $k$ in the random tree $T_{n,\sigma}$ satisfies
\begin{align*}
\mathbb{E}(Y_{n,k})\geq \frac{\gamma n}{c_{\xi}(k+1)\xi(k+1)}-\frac{1}{c_{\xi}\xi(k+1)}.
\end{align*}
\end{lemma}

In order to prove Lemma~\ref{lemma:unbalanced}, the following lemma will be helpful:

\begin{lemma}[\mbox{summation by parts \cite[Thm.~3.41]{Rudin53}}]\label{lemma-summation-by-parts}
Let $m \in \mathbb{N}$ and $a_1, \dots, a_{m}$, $b_1, \dots, b_m$ be real numbers. Then
\begin{align*}
\sum_{i=1}^ma_ib_i = \left(\sum_{i=1}^m a_i\right)b_m + \sum_{i=1}^{m-1}\left(\sum_{\ell = 1}^{i}a_{\ell}\right)\left(b_{i}-b_{i+1}\right).
\end{align*}
\end{lemma}

We are now able to prove Lemma~\ref{lemma:unbalanced}.

\begin{proof}[Proof of Lemma~\ref{lemma:unbalanced}]
As $\xi(x) \in \omega(1/x)$ by condition (a), there is an integer $N_{\xi}$ such that
\begin{equation} \label{eq-xi(k+1)}
\frac{c_{\sigma}-1}{\xi(k+1)} \leq k
\end{equation} 
for all $k \geq N_{\xi}$.
We choose the integer $m$ in such a way that $m \geq \max\{N_{\xi}, N'_{\sigma}, c_{\sigma}\}$ and $\lceil \gamma n \rceil \leq n/2$ for $n >m$.
We prove the statement using induction for $n \geq k+1>m$. For the base case, let $n=k+1$. A binary tree $t \in \mathcal{T}_{k+1}$ has exactly one fringe subtree of size larger than $k$, and we have
\begin{align*}
\mathbb{E}(Y_{k+1,k})= 1 > \frac{\gamma }{c_{\xi}\xi(k+1)}-\frac{1}{c_{\xi}\xi(k+1)}.
\end{align*}
 For the induction step, take an integer $n>k+1$, so that 
\begin{align*}
\mathbb{E}(Y_{i,k})\geq \frac{\gamma i}{c_{\xi}(k+1)\xi(k+1)}-\frac{1}{c_{\xi}\xi(k+1)}
\end{align*}
 for every integer $k+1 \leq i \leq n-1$ by the induction hypothesis. We distinguish two cases.\\

\noindent
\textit{Case $1$:} $\lceil \gamma n \rceil < k+1 \leq n-1$.
We thus have
\begin{align*}
\frac{\gamma n}{c_{\xi}(k+1)\xi(k+1)}-\frac{1}{c_{\xi}\xi(k+1)} < \frac{1}{c_{\xi}\xi(k+1)}-\frac{1}{c_{\xi}\xi(k+1)}=0.
\end{align*}
Furthermore, as $n>k$, we find $\mathbb{E}(Y_{n,k})\geq 1$.
Thus, the statement follows in this case.\\

\noindent
\textit{Case $2$:} $k+1 \leq \lceil \gamma n \rceil$.
As $\lceil \gamma n \rceil \leq n/2$ for $n > m$, we find by Lemma~\ref{lemma:recurrence-expectation}:
\begin{align*}
\mathbb{E}(Y_{n,k})= 1+\sum_{i=k+1}^{n-k-1}\sigma(i,n-i)\left(\mathbb{E}(Y_{i,k})+\mathbb{E}(Y_{n-i,k})\right)+\sum_{i=n-k}^{n-1}\sigma^*(i,n-i)\mathbb{E}(Y_{i,k}).
\end{align*}
By the induction hypothesis, we have
\begin{align*}
\mathbb{E}(Y_{n,k})\geq 1 &+ \sum_{i=k+1}^{n-k-1}\sigma(i,n-i)\left(\frac{\gamma n}{c_{\xi}(k+1)\xi(k+1)}-\frac{2}{c_{\xi}\xi(k+1)}\right) \\
&+ \sum_{i=n-k}^{n-1}\sigma^*(i,n-i)\left(\frac{\gamma i}{c_{\xi}(k+1)\xi(k+1)}-\frac{1}{c_{\xi}\xi(k+1)}\right).
\end{align*}
We introduce the abbreviation
\begin{align*}
S(i)= \frac{\gamma i}{c_{\xi}(k+1)\xi(k+1)}-\frac{1}{c_{\xi}\xi(k+1)}.
\end{align*}
Clearly, $S(i)$ is increasing in $i$. Moreover, set 
\begin{align*}
\sum_{i=k+1}^{n-k-1}\sigma(i,n-i) =: \alpha_1, \quad \sum_{i=n-k}^{n-c_{\sigma}}\sigma^*(i,n-i) =:\alpha_2 \ \text{ and }\sum_{i=n-c_{\sigma}+1}^{n-1}\sigma^*(i,n-i) =:\alpha_3.
\end{align*}
We obtain
\begin{align}\label{eq:lower-bound-equation}
\mathbb{E}(Y_{n,k})
\geq 1 &+ \alpha_1\left(\frac{\gamma n}{c_{\xi}(k+1)\xi(k+1)}-\frac{2}{c_{\xi}\xi(k+1)}\right) 
+ \alpha_3S(n-c_{\sigma}+1)\\
&+ \sum_{i=n-k}^{n-c_{\sigma}}\sigma^*(i,n-i)S(i).\notag
\end{align}
As $\sigma \in \mathcal{L}_{unbal}(\xi, \gamma)$ and additionally $N'_{\sigma} < n$ and $c_{\sigma}< k+1 \leq \lceil \gamma n\rceil$, we have $0 \leq \alpha_1 \leq \xi(k+1)$, $1-\xi(c_{\sigma}) \leq \alpha_3 \leq 1$ and $\alpha_1 + \alpha_2 + \alpha_3 = 1$. 
Furthermore, we have
\begin{align*}
\frac{\gamma n}{c_{\xi}(k+1)\xi(k+1)}-\frac{2}{c_{\xi}\xi(k+1)}\leq S(i) 
\end{align*}
for every integer $i$ with $n-k \leq i \leq n - c_{\sigma}+1$. Hence, we minimize the right-hand side of \eqref{eq:lower-bound-equation} subject to the condition $\alpha_1 + \alpha_2 + \alpha_3 =1$ if we assign the maximal possible weight $\alpha_1=\xi(k+1)$ to the smallest term $\gamma n/(c_{\xi}(k+1)\xi(k+1))-2/(c_{\xi}\xi(k+1))$
and the minimal possible weight $\alpha_3=1-\xi(c_{\sigma})$ to the largest term $S(n-c_{\sigma}+1)$. 
 It remains to bound the sum
\begin{align*}
\sum_{i=n-k}^{n-c_{\sigma}}\sigma^*(i,n-i)S(i)
\end{align*}
from below subject to the condition $\alpha_2 = \xi(c_{\sigma})-\xi(k+1)$. First note that
$\sigma \in \mathcal{L}_{unbal}(\xi, \gamma)$, $N'_{\sigma} < n$ and $\alpha_3 = 1-\xi(c_{\sigma})$
imply
\begin{equation} \label{eq-sum-c_sigma}
\sum_{\ell=c_{\sigma}}^{i}\sigma^*(\ell, n-\ell) =\sum_{\ell=1}^{i}\sigma^*(\ell, n-\ell) -\alpha_3 \geq \xi(c_{\sigma})-\xi(i+1),
\end{equation}
for every $c_{\sigma} \leq i \leq k$ (note that $c_{\sigma}\leq i+1 \leq \lceil \gamma n\rceil$; therefore we can take $r = i+1$ in Definition~\ref{def:unbalanced}).
 Using summation by parts (Lemma~\ref{lemma-summation-by-parts}), we obtain
\begin{align*}
&\sum_{i=n-k}^{n-c_{\sigma}}\sigma^*(i,n-i)S(i) \\
&= \sum_{i=c_{\sigma}}^k\sigma^*(i,n-i)S(n-i)\\
&= \Big(\sum_{i=c_{\sigma}}^k\sigma^*(i,n-i)\Big)S(n-k) 
+ \sum_{i=c_{\sigma}}^{k-1}\Big(\sum_{\ell = c_{\sigma}}^{i}\sigma^*(\ell,n-\ell)\Big)\left(S(n-i)-S(n-i-1)\right)\\
\overset{\text{\eqref{eq-sum-c_sigma}}}&{\geq} \left(\xi(c_{\sigma})-\xi(k+1)\right)S(n-k)+\sum_{i=c_{\sigma}}^{k-1}\left(\xi(c_{\sigma})-\xi(i+1)\right)\left(S(n-i)-S(n-i-1)\right)\\
&= -\xi(k+1)S(n-k) +\xi(c_{\sigma})S(n-c_{\sigma})- \sum_{i=c_{\sigma}}^{k-1}\xi(i+1)\left(S(n-i)-S(n-i-1)\right).
\end{align*}
Note that we have
\begin{equation*} 
S(n-i)-S(n-i-1) = \frac{\gamma}{c_{\xi}(k+1)\xi(k+1)}.
\end{equation*}
 Altogether, we obtain from \eqref{eq:lower-bound-equation} that
\begin{align*}
\mathbb{E}(Y_{n,k}) & \geq 1+\xi(k+1)\left(\frac{\gamma n}{c_{\xi}(k+1)\xi(k+1)}-\frac{2}{c_{\xi}\xi(k+1)}\right)\\
&\qquad + \, \frac{\gamma n}{c_{\xi}(k+1)\xi(k+1)} - \frac{\gamma (c_{\sigma}-1)}{c_{\xi}(k+1)\xi(k+1)}-\frac{1}{c_{\xi}\xi(k+1)}\\
&\qquad  - \, \xi(k+1)\left(\frac{\gamma n}{c_{\xi}(k+1)\xi(k+1)}-\frac{\gamma k}{c_{\xi}(k+1)\xi(k+1)}-\frac{1}{c_{\xi}\xi(k+1)}\right)\\
&\qquad   - \sum_{i=c_{\sigma}}^k\frac{\xi(i)\gamma}{c_{\xi}(k+1)\xi(k+1)}\\
&= 1 - \frac{1}{c_{\xi}}+\frac{\gamma n}{c_{\xi}(k+1)\xi(k+1)}-\frac{\gamma (c_{\sigma}-1)}{c_{\xi}(k+1)\xi(k+1)}-\frac{1}{c_{\xi}\xi(k+1)}\\
&\qquad  + \frac{\gamma k}{c_{\xi}(k+1)}-\sum_{i=c_{\sigma}}^k\frac{\xi(i)\gamma}{c_{\xi}(k+1)\xi(k+1)}.
\end{align*}
Since we have
\begin{align*}
\sum_{i=c_{\sigma}}^k\xi(i)\leq c_{\xi}(k+1)\xi(k+1)
\end{align*}
by the assumption from condition (b), we get
\begin{align*}
\mathbb{E}(Y_{n,k}) \geq 1&-\frac{1}{c_{\xi}}+\frac{\gamma n}{c_{\xi}(k+1)\xi(k+1)}-\frac{\gamma (c_{\sigma}-1)}{c_{\xi}(k+1)\xi(k+1)}-\frac{1}{c_{\xi}\xi(k+1)}\\
 &+\frac{\gamma k}{c_{\xi}(k+1)}-\gamma.
\end{align*}
Furthermore, from  \eqref{eq-xi(k+1)} we obtain
\begin{align*}
-\frac{\gamma (c_{\sigma}-1)}{c_{\xi}(k+1)\xi(k+1)}+\frac{\gamma k}{c_{\xi}(k+1)} \geq 0
\end{align*}
and hence
\begin{align*}
\mathbb{E}(Y_{n,k}) \geq 1&-\frac{1}{c_{\xi}}+\frac{\gamma n}{c_{\xi}(k+1)\xi(k+1)}-\frac{1}{c_{\xi}\xi(k+1)}-\gamma.
\end{align*}
Finally, as $1/c_{\xi}+\gamma \leq 1$ by assumption (by condition (b)), the statement follows.
\end{proof}

We are now able to prove Theorem~\ref{thm:lower-bound-unbalanced}. The main idea of the proof is based on techniques from \cite{RalaivaosaonaW15, SeelbachW22}.

\begin{proof}[Proof of Theorem~\ref{thm:lower-bound-unbalanced}]
Let $\delta_n$ for $n \in \mathbb{N}$ be defined by
$
\delta_n=2\ell_{\sigma}\log \log n /\log n,
$
where $\ell_{\sigma}\geq 2$ is the constant from condition (c) in Theorem~\ref{thm:lower-bound-unbalanced}. 
Moreover, let 
$$k_1 =\left\lceil N_{\sigma} \bigg( 1 + \frac{(1+\delta_n)\log n}{\log(1/\rho)} \bigg) \right\rceil.$$ 
Furthermore, assume that $n$ is sufficiently large, so that the lower bound on $\mathbb{E}(Y_{n,k})$ from Lemma~\ref{lemma:unbalanced} applies for $k \ge k_1-1$, and that $k_1 < (\log n)^{\ell_{\sigma}}$.
The random number $F_{n,\sigma}$ of distinct fringe subtrees in the random tree $T_{n,\sigma}$ is bounded below by the number of distinct fringe subtrees in $T_{n,\sigma}$ of sizes $k$ for $k_1\leq k \leq n^{\delta_n/2}$, where $n^{\delta_n/2}=(\log n)^{\ell_{\sigma}}$ by definition of $\delta_n$. Let $X_{n,k}$ denote the (random) number of fringe subtrees of size exactly $k$ occurring in $T_{n,\sigma}$, and let $X_{n,k}^{(2)}$ denote the (random) number of unordered pairs of identical fringe subtrees of size exactly $k$ occurring in a random tree $T_{n,\sigma}$. Using the inclusion-exclusion principle, we find that $\mathbb{E}(F_{n,\sigma})$ can be bounded from below as follows:
\begin{align}\label{eq:lower-bound-siebformel}
\mathbb{E}(F_{n,\sigma})\geq \sum_{k_1 \leq k \leq n^{\delta_n/2}}\mathbb{E}( X_{n,k}) - \sum_{k_1 \leq k \leq n^{\delta_n/2}} \mathbb{E}(X_{n,k}^{(2)}).
\end{align}

We first bound the second sum on the right-hand side of \eqref{eq:lower-bound-siebformel}. By Lemma~\ref{lemma:lower-bound-leafsize}, we find that the number of fringe subtrees of size at least $N_{\sigma}+1$ in a binary tree $t$ of size $k>N_{\sigma}$ is at least $k/N_{\sigma}-1$. As $\sigma \in \mathcal{L}_{\text{up}}(\rho)$ with integer $N_{\sigma}$, we find
\begin{align*}
P_{\sigma}(t)=\prod_{v \text{ inner node of $t$}}\sigma(|t_{\ell}[v]|, |t_r[v]|)\leq \prod_{v \text{ node of $t$}\atop |t[v]|>N_{\sigma}}\sigma(|t_{\ell}[v]|, |t_r[v]|)\leq \rho^{k/N_{\sigma}-1}
\end{align*}
for every binary tree $t \in \mathcal{T}_k$.
Let us condition on the event that $X_{n,k}=N$ for some natural number $N$. These $N$ fringe subtrees are independent random trees of size $k$, and the probability that such a fringe subtree equals a given binary tree $t \in \mathcal{T}_k$ is given by $P_{\sigma}(t)$. Thus, we have
\begin{align*}
\mathbb{E}(X_{n,k}^{(2)} \mid X_{n,k}=N)=\binom{N}{2}\sum_{t \in \mathcal{T}_k}P_{\sigma}(t)^2 \leq n^2\rho^{k/N_{\sigma}-1}\sum_{t \in \mathcal{T}_k}P_{\sigma}(t)=n^2\rho^{k/N_{\sigma}-1}.
\end{align*}
Since this upper bound on $\mathbb{E}(X_{n,k}^{(2)} \mid X_{n,k}=N)$ holds independently of $N$, we obtain with the law of total expectation
\begin{align*}
\mathbb{E}(X_{n,k}^{(2)})=\sum_{N=0}^n\mathbb{E}(X_{n,k}^{(2)} \mid X_{n,k}=N)\mathbb{P}(X_{n,k}=N)\leq n^2\rho^{k/N_{\sigma}-1}.
\end{align*}
Since $k_1 \geq N_{\sigma} \big( 1 + \frac{(1+\delta_n)\log n}{\log(1/\rho)} \big)$, we obtain for all $k \geq k_1$:
\begin{align*}
\mathbb{E}(X_{n,k}^{(2)})\leq n^{1-\delta_n}.
\end{align*}
We thus have
\begin{align*}
\sum_{k_1 \leq  k \leq n^{\delta_n/2}}\mathbb{E}(X_{n,k}^{(2)})\leq n^{1-\delta_n/2}=\frac{n}{(\log n)^{\ell_{\sigma}}} \in o\left(\frac{n}{\log n}\right).
\end{align*}
It remains to bound the first sum on the right-hand side of \eqref{eq:lower-bound-siebformel}:
If $Y_{n,k}$ again denotes the random number of fringe subtrees of size larger than $k$ in the random tree $T_{n,\sigma}$, we have
\begin{align*}
\sum_{k_1 \leq  k \leq n^{\delta_n/2}}\mathbb{E}(X_{n,k})=\mathbb{E}(Y_{n,k_1-1})-\mathbb{E}(Y_{n,n^{\delta_n/2}}).
\end{align*}
By condition (c) of Theorem~\ref{thm:lower-bound-unbalanced}, we have $\mathbb{E}(Y_{n,n^{\delta_n/2}}) \in o(n/\log n)$. Moreover, by Lemma~\ref{lemma:unbalanced}, we find that 
\begin{align*}
\mathbb{E}(Y_{n,k_1-1})\geq \frac{\gamma n}{c_{\xi}k_1\xi(k_1)}-\frac{1}{c_{\xi}\xi(k_1)} \in \Omega\left(\frac{n}{\log n}\right),
\end{align*}
since $\xi(x) \leq 1$.
By monotonicity of $\xi$, we have $\xi(k_1)\leq \xi\big(\frac{N_{\sigma}\log n}{\log(1/\rho)}\big)$, so we obtain from \eqref{eq:lower-bound-siebformel} that
\begin{align*}
\mathbb{E}(F_{n,\sigma})\geq \frac{\gamma \log(1/\rho) n}{c_{\xi}N_{\sigma}\xi\big(\frac{N_{\sigma}\log n}{\log(1/\rho)}\big)\log n}\left(1+o(1)\right).
\end{align*}
This finishes the proof.
\end{proof}

\begin{example}\label{ex:uniform-unbalanced}
In this example, we use Theorem~\ref{thm:lower-bound-unbalanced} in order to prove a lower bound on the expected number of distinct fringe subtrees in a random tree $T_{n,\sigma_{\text{uni}}}$, i.e., in a random tree of size $n$ drawn from the set $\mathcal{T}_n$ according to the uniform probability distribution $\sigma_{\text{uni}}$ from Example~\ref{ex:uniform}.
We have to show that all requirements of Theorem~\ref{thm:lower-bound-unbalanced} are satisfied. Let us write $\sigma$ for  $\sigma_{\text{uni}}$ in the following.
Recall that 
$$
\sigma(i,n-i) = \frac{C_{i-1} C_{n-i-1}}{C_{n-1}}.
$$
(i) $\sigma \in \mathcal{L}_{\text{up}}(\rho)$ for a constant $\rho \in (0,1)$.
A short computation shows
that for a fixed integer $n$, the maximal value of $\sigma(i,n-i)$ with $1 \leq i \leq n-1$ is attained at $i_{max}(n)=1$ and $i_{max}'(n)=n-1$. In particular, we have 
\begin{align*}
\sigma(i,n-i)\leq \sigma(1,n-1)\leq \frac{n}{4n-6}
\end{align*} 
for all $1 \leq i \leq n-1$. If we fix $N_\sigma \in \mathbb{N}$, we thus find that $\sigma \in \mathcal{L}_{\text{up}}(\rho)$ with integer $N_\sigma$, where $\rho=N_\sigma/(4N_\sigma-6)$. For the sake of simplicity, let $N_{\sigma}=3$. Then $\sigma \in \mathcal{L}_{\text{up}}(\rho)$ with $\rho=1/2$.
\\

\noindent
 (ii) $\sigma \in \mathcal{L}_{unbal}(\xi, \gamma)$ for a decreasing function $\xi$ and a constant $\gamma$.
Let $0<\gamma < 1/2$, and let $\varepsilon_0>0$ and $\delta_0 >1$. From the asymptotic formula for the Catalan numbers \eqref{eq:Catalan-growth}, we find that there is an integer $N_{\sigma}'$ (depending on $\varepsilon_0$ and $\delta_0$) such that both
\begin{align*}
\frac{(1-\varepsilon_0)4^n}{\sqrt{\pi}n^{3/2}}\leq C_n \leq \frac{(1+\varepsilon_0)4^n}{\sqrt{\pi}n^{3/2}}
\end{align*}
and 
\begin{align*}
\frac{n^{5/2}}{(n-2)^2\sqrt{(r-2)(n-\gamma n -1)}}\leq   \frac{\delta_0}{\sqrt{(1-\gamma)r}}
\end{align*}
for all $n,r \geq N_{\sigma}'$. Set $c_{\sigma}=N_{\sigma}'$, and suppose that $c_{\sigma} \leq r \leq \lceil \gamma n \rceil$. Then
\begin{align*}
\sum_{r \leq i \leq n-r}\sigma(i,n-i)&=\sum_{r \leq i \leq n-r}\frac{C_{i-1}C_{n-i-1}}{C_{n-1}}\\
&\leq \frac{(1+\varepsilon_0)^2(n-1)^{3/2}}{(1-\varepsilon_0)4	\sqrt{\pi}}\sum_{r \leq i \leq n-r}(i-1)^{-3/2}(n-i-1)^{-3/2}\\
& \leq \frac{(1+\varepsilon_0)^2n^{3/2}}{(1-\varepsilon_0)2	\sqrt{\pi}}\int_{r-1}^{n/2}(x-1)^{-3/2}(n-x-1)^{-3/2}\mathrm{d}x\\
&= \frac{(1+\varepsilon_0)^2n^{3/2}}{(1-\varepsilon_0)	\sqrt{\pi}}\frac{n-2r+2}{(n-2)^2\sqrt{(r-2)(n-r)}}\\
&\leq \frac{(1+\varepsilon_0)^2}{(1-\varepsilon_0)	\sqrt{\pi}}\frac{n^{5/2}}{(n-2)^2\sqrt{(r-2)(n-\gamma n - 1)}}\\
&\leq \frac{(1+\varepsilon_0)^2}{(1-\varepsilon_0)	\sqrt{\pi}}\frac{\delta_0}{\sqrt{(1-\gamma)r}}.
\end{align*}
Set
\begin{align*}
\delta_1:=\frac{(1+\varepsilon_0)^2\delta_0}{1-\varepsilon_0}>1 \quad\text{ and } \quad \xi(r):=\frac{\delta_1}{\sqrt{\pi(1-\gamma)r}}.
\end{align*}
Then $\sigma \in \mathcal{L}_{unbal}(\xi, \gamma)$, where the constant $\delta_1>1$ can take values arbitrarily close to $1$ for suitable choices of $\varepsilon_0$ and $\delta_0$. \\

\noindent
(iii) Condition (a) of Theorem~\ref{thm:lower-bound-unbalanced} is clearly satisfied, i.e., $\xi_{uni}(x) \in \omega(1/x)$.\\

\noindent
(iv) Condition (b) is satisfied as well:
We have
\begin{align*}
\sum_{i=N_{\sigma}'}^k\xi(i) &=\frac{\delta_1}{\sqrt{\pi(1-\gamma)}}\sum_{i=N_{\sigma}'}^ki^{-1/2}
\leq \frac{\delta_1}{\sqrt{\pi(1-\gamma)}} \int_{N_{\sigma}'-1}^{k} x^{-1/2}\mathrm{d}x\\
&\leq  \frac{2\delta_1 \sqrt{k+1} }{\sqrt{\pi(1-\gamma)}}=2(k+1)\xi(k+1).
\end{align*}
Thus, condition (b) holds with $c_{\xi_{uni}}=2$ (as $\gamma < 1/2$, we also have $1/c_{\xi_{uni}}+\gamma \leq 1$).\\

\noindent
(v)  Finally, condition (c) of Theorem~\ref{thm:lower-bound-unbalanced} holds as well:
From Example~\ref{ex:uniform-strongly-balanced}, we know that $\sigma \in \mathcal{L}_{\text{sbal}}(\vartheta,\gamma)$ with $\vartheta(x) \in \Theta(x^{-1/2})$ for $0 < \gamma < 1/2$. If we set $\ell_{\sigma}=4$, then
Lemma~\ref{lemma:strongly-balanced} yields
\begin{align*}
\mathbb{E}(Y_{n,(\log n)^4}) \in o\left(\frac{n}{\log n}\right).
\end{align*}
We can therefore apply Theorem~\ref{thm:lower-bound-unbalanced}. The expected number of distinct fringe subtrees in a random tree of size $n$ drawn according to the uniform probability distribution satisfies
\begin{align*}
\mathbb{E}(F_{n,\sigma_{\text{uni}}})&\geq \frac{\gamma n \log(1/\rho_{uni})}{c_{\xi}N_{\sigma}\xi\big(\frac{N_{\sigma}\log n}{\log(1/\rho)}\big)\log n}(1+o(1))\\
&=\frac{\gamma \sqrt{(1-\gamma)\pi\log(1/\rho)}}{\delta_1c_{\xi}\sqrt{N_{\sigma}}}\cdot \frac{n}{\sqrt{\log n}}(1+o(1))\\
&=\frac{\gamma \sqrt{(1-\gamma)\pi}}{2\sqrt{3}\delta_1}\cdot \frac{n}{\sqrt{\log n}}(1+o(1)).
\end{align*}
As $0 < \gamma < 1/2$ is arbitrary, we can choose $\gamma$ arbitrarily close to the optimal value $1/2$.
We then obtain for $\sigma = \sigma_{\text{uni}}$
\begin{equation*}
\mathbb{E}(F_{n,\sigma_{\text{uni}}})
\geq 
\frac{ \sqrt{\pi}}{4\sqrt{6}\delta_1}\cdot \frac{n}{\sqrt{\log n}}(1+o(1))
\end{equation*}
for any constant $\delta_1 > 1$. Recall that in Example~\ref{ex:uniform-strongly-balanced}, we have already shown that
\begin{align*}
\mathbb{E}(F_{n,\sigma_{\text{uni}}}) \leq 8\delta\sqrt{2\pi}\cdot \frac{n}{\sqrt{\log n}}(1+o(1)),
\end{align*}
for any constant $\delta>1$. In particular, we thus have $\mathbb{E}(F_{n,\sigma_{\text{uni}}}) \in \Theta(n/\sqrt{\log n})$.
We remark again that in \cite{FlajoletSS90,RalaivaosaonaW15,SeelbachW22}, it is shown that in fact
$$\mathbb{E}(F_{n,\sigma_{\text{uni}}})=\frac{2\sqrt{2}}{\sqrt{\pi}}\cdot \frac{n}{\sqrt{\log (n)}}(1+o(1)).$$ 
\end{example}

It remains to remark that Theorem~\ref{thm:lower-bound-unbalanced} can be applied to a wide variety of leaf-centric binary tree sources, including the binary search tree model from Example~\ref{ex:bst} and the binomial random tree model from Example~\ref{ex:dst}: For these two leaf-centric binary tree sources, Corollary \ref{cor:unbalanced} again yields asymptotic lower bounds of the form $\mathbb{E}(F_{n,\sigma_{\text{bst}}}) \in \Omega(n/\log n)$ and $\mathbb{E}(F_{n,\sigma_{\text{bin},p}}) \in \Omega(n/\log n)$, respectively.

\section{Conclusion and Open Problems}

We have proposed several classes of leaf-centric binary tree sources and derived upper and lower bounds on the number of distinct fringe subtrees occurring in a random tree generated by a leaf-centric binary tree source from the respective class. 
For the class of critical $\beta$-splitting random trees, it remains open to determine the asymptotic growth of the average
number of distinct fringe subtrees.

Another type of binary tree sources are \emph{depth-centric binary tree sources} \cite{GanardiHLS19, ZhangYK14}, which yield probability distributions on the set of binary trees of a fixed depth and resemble leaf-centric tree sources in many ways. Furthermore, leaf-centric binary tree sources have been generalized to random tree models for plane trees called \emph{fixed-size ordinal tree sources}
in \cite{MunroNSW21}. 

An interesting problem would be to estimate the number of distinct fringe subtrees with respect to classes of depth-centric or fixed-size ordinal tree sources.
Furthermore, another open problem is to consider the question of estimating the number of distinct fringe subtrees in a leaf-centric binary tree source under a generalized interpretation of ``distinctness'', as was done in \cite{SeelbachW22} for simply generated families of trees and families of increasing trees. 

\end{document}